\documentclass[10pt,
aps,pra,
reprint,
notitlepage,
superscriptaddress,
frontmatterverbose, 
showpacs,
amsmath,amssymb
]{revtex4-1}
    \makeatletter
    \let\@fnsymbol\@alph
    \makeatother
\usepackage[utf8]{inputenc}
\usepackage[version=3]{mhchem} 
\usepackage{siunitx}
\usepackage{nicefrac}
\usepackage{graphicx}
\usepackage[british]{babel}
\usepackage{ifpdf}
\usepackage{natbib}
\usepackage{multirow}
\usepackage{appendix}
\usepackage{hyperref}

\hypersetup{pdfauthor={Salvador Rodríguez Gómez},pdftitle={The Si-Ge substitutional series in the chiral STW Zeolite Structure Type}}
\begin{document}
\bibliographystyle{unsrtnat}
\preprint{}
\title{The Si-Ge substitutional series in the chiral STW Zeolite Structure Type}
\author{Reus T. \surname{Rigo}}
\thanks{\textbf{These two authors contributed equally to this work.}}
\affiliation{Instituto de Ciencia de Materiales de Madrid (ICMM), Consejo Superior de Investigaciones Cient{\'i}ficas (CSIC), Sor Juana In{\'e}s de la Cruz 3, 28049 Madrid, Spain.}
\author{Salvador \surname{R. G. Balestra} }
\thanks{\textbf{These two authors contributed equally to this work.}}
\affiliation{Departament of Physical, Chemical and natural Systems, Universidad Pablo de Olavide, Ctra. Utrera km 1, 41013 Seville, Spain}
\author{Said \surname{Hamad}}
\affiliation{Departament of Physical, Chemical and natural Systems, Universidad Pablo de Olavide, Ctra. Utrera km 1, 41013 Seville, Spain}
\author{Rocío \surname{Bueno-Pérez}}
\affiliation{Departament of Physical, Chemical and natural Systems, Universidad Pablo de Olavide, Ctra. Utrera km 1, 41013 Seville, Spain}
\author{A. Rabdel \surname{Ru\'{i}z-Salvador}}
\email[Corresponding autor:]{rruisal@upo.es}
\affiliation{Departament of Physical, Chemical and natural Systems, Universidad Pablo de Olavide, Ctra. Utrera km 1, 41013 Seville, Spain}
\author{Sof\'{i}a \surname{Calero}}
\email[Corresponding autor:]{scalero@upo.es}
\affiliation{Departament of Physical, Chemical and natural Systems, Universidad Pablo de Olavide, Ctra. Utrera km 1, 41013 Seville, Spain}
\author{Miguel A. \surname{Camblor}}
\email[Corresponding autor:]{macamblor@icmm.csic.es}
\affiliation{Instituto de Ciencia de Materiales de Madrid (ICMM), Consejo Superior de Investigaciones Cient{\'i}ficas (CSIC), Sor Juana In{\'e}s de la Cruz 3, 28049 Madrid, Spain.}
\date{\today}
\begin{abstract}
The whole  compositional range (Ge$_f$ = Ge/(Ge+Si)= 0 to 1) of zeolite STW has been synthesized and studied by a comprehensive combined experimental--theoretical approach. The yield of zeolite goes through a maximum and then drops at the \ce{GeO2} side of the series, following the inverse of the calculated free energy curve. The unit cell generally expands, roughly linearly, as the Ge$_f$ increases, but a notable resilience to expansion is observed at the high silica side. This can be attributed to the more rigid character of \ce{SiO2} and the ability of Ge units to deform. Density functional theory calculations provide a new assignment of the previously controversial $^{19}$F MAS NMR resonances for occluded fluoride, which is based not only in the number of Ge atoms in the double-4-ring units but also on the way they are associated (namely, no Ge, isolated Ge, Ge pairs or closed Ge clusters). While we found an overall good agreement between the experimental and theoretical trends in preferential occupation by Ge of different crystallographic sites, the theoretical models show more sharp and abrupt tendencies, likely due both to limitations of the approach and to kinetic factors that allow metastable configurations to actually exist.

\end{abstract}
\maketitle
\section{Introduction}

Zeolites find an extraordinarily wide commercial applicability,\cite{Sherman2000} and this in turn fosters further research aimed to the synthesis of zeolites with new structures and compositions.\cite{Cundy2003}  Among the many factors determining the phase that crystallizes in a zeolite synthesis,\cite{Lu2018a} the organic structure directing agents (SDA),\cite{Davis1992} fluoride anions,\cite{Caullet2005} and framework elements other than Si and Al (Ge, Zn, Be, Ga \dots) may afford the discovery of new zeolite structures.\cite{Camblor2011} In particular, Germanium, specially when used together with fluoride, tend to produce structures with double 4-ring  units (D4R).\cite{Conradsson2000,Corma2001,Villaescusa2003} Despite the low stability of Ge-zeolites upon both calcination,\cite{Villaescusa2016a} and hydrolysis by ambient moisture of the calcined materials,\cite{Shamzhy2016} the discovery of new Zeolite Framework Types (ZFT),\cite{Baerlocher2004} even if unstable, is still of interest. In fact, the weakness of Ge-zeolites has been advantageously used to derive new materials from them through the assembly-disassembly-organisation-reassembly strategy (ADOR), which has so far produced several interesting zeolites that are, in addition, more stable than the parent one.\cite{Roth2013,Eliasova2015,Wheatley2014,Chlubna2014} These derived zeolites may be 'unfeasible' to obtain by the conventional hydrothermal routes, adding interest to Ge-zeolites.\cite{Mazur2016} On the other hand, Ge-zeolites may be stabilized by postsynthetic treatments by substituting Ge by Si or Al.\cite{Gao2009,Burel2014,Xu2014d,Shamzhy2016} Finally, unstable but structurally interesting zeolites discovered by using Ge, such as the chiral zeolite STW,\cite{Tang2008} can become a target for the synthesis of more stable materials with the same structure, as was the case for the STW pure silica version, HPM-1.\cite{Rojas2012a,Schmidt2014} 

STW was first realized as a germanosilicate.\cite{Tang2008} Its interest relies on its chiral nature and the presence of a helicoidal medium pore channel. Every single crystal is homochiral but standard synthesis procedures using achiral organic SDA are expected to yield racemic conglomerates.\cite{Tang2008,Rojas2012a,Schmidt2014} However, very recently it has been possible to prepare enantiomerically enriched scalemic conglomerates by using an enantiomerically pure chiral dication, and the materials proved to yield small but significant enantiomeric excess in both asymmetric catalysis and adsorption processes.\cite{Brand2017} These syntheses produced germanosilicate and aluminogermanosilicates, but recent studies suggest homochiral STW silica phases may as well be possible.\cite{Lu2018} These silica zeolites are expected to be not only much more stable but also more amenable to selective separations, since the larger flexibility of \ce{GeO2} frameworks appears to be detrimental to chiral recognition.\cite{Bueno-Perez2018} Here we report that substitution of Si by Ge in the chiral D4R-containing zeolite structure STW can be attained for any value of the Ge molar fraction (Ge$_f$=Ge/(Ge+Si)). By combining experiment and theory we have been able to get significant insight into that system, particularly on the energetics of the zeolite, the unit cell expansion, which is buffered at the low Ge$_f$ side, the previously controversial assignment of the fewer than expected $^{19}$F MAS NMR resonances, and the differential occupation of crystallographic sites as Ge$_f$ increases. This insight is expected to be of general interest within the field of Ge-zeolites and their flourishing derived strategies to develop new materials.\cite{Roth2013,Eliasova2015,Wheatley2014,Chlubna2014,Mazur2016}

\section{Experimental}

\subsection{Synthesis}
All the zeolite syntheses were done using equimolar amounts of hydrofluoric acid and 2-ethyl-1,3,4-trimethylimidazolium (2E134TMI) hydroxide. 2E134TMI was synthesized as iodide salt and exchanged to the hydroxide form as previously reported.\cite{Rojas2013b} The synthesis mixture was prepared by adding (if required) first tetraethylorthosilicate (TEOS, 98\% Aldrich) and then (if required) germanium dioxide (99.998\% Aldrich) to a concentrated solution of 2E134TMI hydroxide. The mixture was stirred at room temperature allowing evaporation of ethanol (if TEOS was used) and water, until the desired composition was reached. Evaporation was monitored by weight. Then, hydrofluoric acid (48 wt\%, Sigma--Aldrich) was added to the gel and stirred with a spatula for approximately 15 minutes. The obtained gel was transferred to Teflon vessels inside stainless steel autoclaves, which were heated in an oven at a temperature of 175 $^\circ$C while tumbling at 60 rpm. At preselected times (generally close to 24, 48, 144 and 240 hours), the autoclaves were removed from the oven and quenched and the product filtered on paper or centrifuged, washed with deionized water and dried at 100 $^\circ$C. The final composition of the gel was: $(1-x)$\ce{SiO2} : $x$\ce{GeO2} : $0.5$ 2E134TMIOH : $0.5$ HF : $4$\ce{H2O}, where $x = \text{Ge}/(\text{Si}+\text{Ge})$ is the molar fraction of germanium oxide, which will be expressed in the following as Ge$_f$. 

\subsection{Characterization} 
Power X-ray diffraction was performed using a Bruker D8 Advance diffractometer, with Cu $K_\alpha$ radiation in the 3.5--45 $^\circ$ $2\theta$ range. The unit cell of HPM-1 samples with varying Ge$_f$ were refined by a least squares regression procedure using the program UnitCell and 16 reflections uniquely indexed in space group P6$_1$22, covering the 8--30 $^\circ$ 2$\theta$ range.\cite{Holland1997} Synchrotron X-ray powder diffraction data were collected at the SpLine BM25A at the ESRF, Grenoble, in capillary mode (0.8 mm diameter) using monochromatic radiation ($\lambda$ = 0.56383 \AA) for the samples synthesized with Ge$_f$=0.4, 0.6 and 1.0. Rietveld refinement was performed using GSAS,\cite{Larson2004} under the EXPGUI graphical interface.\cite{Toby2001} C,N,H analyses were performed with a LECO CHNS-932 instrument. Ge and Si chemical analysis were performed by Inductively Couple Plasma--Mass Spectrometry (ICP-MS) using an ICP-MS NexION 300XX equipment. $^{19}$F, $^{29}$Si, $^{1}$H and $^{13}$C MAS NMR experiments were recorded on a Bruker AV 400WB, as described elsewhere.\cite{Rojas2012} Field emission scanning electron microscopy (FE-SEM) images were obtained with a FEI NOVA NANOSEM 230 without metal coating. Thermogravimetric analyses were obtained with an SDT Q600 from TA Instruments at a heating rate 10 $^\circ$ C min$^{-1}$ under an air flow of 100 mL/min.  

\subsection{Theoretical methodology}
The STW framework has a large unit cell with sixty tetrahedral sites with five symmetrically distinct sites.\cite{Rojas2013b} This conduces to massive amounts of symmetrically unique Si-Ge configurations for each Ge content by unit cell in the interval from 4 to 56. To deal with this, we employ a recently developed method that evaluates the energy of the system using an effective Hamiltonian (EH, see Supplementary Information).\cite{Arce2018,MottLittentonCatlow} The EH was parameterized using the lattice energy of the pure \ce{SiO2} STW zeolite and the substitution energies for 1 to 4 Ge atoms by unit cell computed with interatomic potentials. Explicit calculations of the lattice energy are computed for the whole set of possible symmetry non-redundant configurations up to 3 Ge by unit cells. The selection of these configurations was achieved with the Site Occupancy Disorder (SOD) program,\cite{GrauCrespo2007} which allows to reduce at least by one order of magnitude the computational cost by discarding the redundant configurations. Details of the number of configurations used for each Ge content are collected in the Supplementary Information (see Table \ref{tbl:number_of_conf}).

For four Ge per unit cell, 40890 inequivalent configurations appear, which represented a heavy computational cost. We noted during the energy minimization with lower Ge content that the convergence of the calculations is rather slow, as compared to aluminosilicate zeolites due to the presence of multiple local minima on the total potential energy surface, which we identify as originated by the larger flexibility of the solids associated to the presence of Ge. Therefore, the lattice energy of ca. 3.68 \% of the 4-Ge configurations were explicitly computed by standard atomistic methods and the remaining 96.31 \% by using the EH. The selection of this 3.68 \% was achieved by considering those relevant configurations having three neighbour Ge atoms and the fourth one as first or second neighbour of one of those three. A more spread distribution of Ge atoms causes a lower effect on  the local structure and can therefore be accurately described by the EH. Once the EH was parameterized, it was used to compute the lattice energy of each configuration (see Equation S7). The atomistic calculations were performed with the GULP code,\cite{JD1997} using \citeauthor{Sastre2003} interatomic potentials,\cite{Sastre2003} which have been used in the past to predict preferable occupation of Ge in D4R.\cite{Pulido2006,Pulido2006b} Short-range Buckingham potential was evaluated within a cut-off of 16 \AA, while the long-range Coulomb potential was calculated by the Ewald method.\cite{ewald1921berechnung} Energy minimization was performed with the BFGS minimizer,\cite{Press:2007:NRE:1403886} switching to RFO method after a suitable progress of the structural relaxation to remove the existing imaginary vibrational modes, if any, and therefore providing true energy minima structures. This procedure has been proven to be particularly useful for modelling zeolitic materials.\cite{AlmoraBarrios2001,Lewis2002,Balestra2015}

To reduce the size-effect contribution in the error of averaged observables we have designed an ensemble of special quasirandom structures (SQS's),\cite{Zunger1990} that mimic the average in composition of the calculated structures and radial correlation functions of optimised structures for each molar fraction. The generation of these structures take into account the free energy of the unit cell for each Ge content and correlation functions. SQS's have been extensively used in substitutionally random A$_x$B$_{1-x}$ solids in the past,\cite{Zunger1990,Urban2016,Saltas2017} but never in nanoporous crystals, to our knowledge.

The use of the EH and SQS's allowed us to evaluate the free energy of formation of the zeolites in the complete range of Si/Ge content, by appropriate Boltzmann weighing and considering also the configurational entropic contribution. For each given number of Ge atoms per unit cell, we selected the 50 lowest energy configurations for the theoretical estimation of the structural features. They were subject to interatomic potentials full lattice energy minimizations using the same type of calculations described above. Since we are interested in understanding the behaviour of the cell parameters and volume as a function of the Ge molar fraction and the size constrains of our calculations leads to small variations of the hexagonal symmetry, we renormalized $a$ and $b$ parameters. For this, we take for each configuration the cell volume invariant given by the energy minimization, as well as the cell parameters and force the cell angles to be $\alpha=\beta=90^\circ$ and $\gamma=120^\circ$, while recalculating $a$ and $b$ cell parameters.

An open source \texttt{Fortran 2003} code for the manipulation and generation of the most probable structures (using the effective Hamiltonian analysis and the special quasirandom structures theory) is available at \url{https://github.com/salrodgom/ising_cation_3D_nanoporous}.

Density Functional Theory calculations were conducted to compute the $^{19}$F-NMR chemical shift, using the linear response method.\cite{Pickard2001,Yates2007} It is know that calculations of chemical shifts computed using the PBE functional,\cite{PhysRevLett.77.3865} might differ from the experimental measured values in cases where the covalency of the system is significant,\cite{Profeta2004,Sadoc2011,Pedone2012,Laskowski2012} with differences as high as 80 ppm.\cite{Zheng2009,Martineau2016} Due to the large size of our system, we cannot perform the types of the electronic structure calculations with hybrid DFT functionals that would solve this issue. We then have to use PBE, in spite of quantitative agreement between experimental and computed chemical shifts is not expected, as we have a system with varying degree of covalency of the Ge--F interactions at the different SiGe-F environments. However, we do expect to use the computed values to identify the existence of distinct resonance peaks. Due to the large number of Ge substitutions that we will study, a wide range of resonance peaks is expected to appear, and we will try to assign the experimentally observed peaks using the theoretical chemical shifts.

Regarding the generation of the Ge-substituted zeolite for the NMR calculations, as mentioned above, the number of possible non-equivalent configurations is too high. Besides, the geometry optimization of each configuration is computationally expensive, due to the combination of a large size of the zeolite, Pulay forces (inherent to the plane wave DFT calculation of periodic solids with varying cell volume), and the large structural flexibility caused both by the zeolite framework and the presence of extra-framework species (fluor counteranions and template cations).
As a consequence, we manually created eight zeolite structures, which allow us to include all possible Si-Ge distributions (about 20) within the D4R units. Since STW zeolite has 6 D4R cubes by unit cell, by constructing eight zeolite structures we ensure that each Si-Ge D4R distribution was considered at least two times, which is useful to validate transferability of the results and to increase the size of the sampling space to provide more accurate estimation of the errors.
Several works report the calculation of the chemical shift using non-periodic, isolated D4R units, having the F atom inside. It is assumed then that the chemical shift is a local property, which depends only on the composition and Si-Ge distribution of a given D4R cube, and on the resulting chemical environment of the F atom. Details of these structures are provided in the supplementary information.

All DFT calculations were performed with the VASP program,\cite{PhysRevB.47.558, Kresse1994,PhysRevB.54.11169,Kresse1996} using the PAW potentials,\cite{PhysRevB.59.1758} the PBE functional connected to the D3 van der Waals potential,\cite{Grimme2010} and 900 eV cut-off for both the geometry optimization and the NMR data calculation. Since the calculation of chemical shifts requires the computation of magnetic energy levels separated by very small gaps, very tight minimization criteria for the structural relaxations are needed. The calculations of the chemical shift on STW structures optimized using 600 eV cut-off did not converge in six out of eight structures, which caused the need to perform very expensive energy optimizations, with 900 eV cut-off. All calculations were performed with the gamma point only, due to the large size of the unit cell.

\section{Results and discussion}
\subsection{Synthesis}

The use of 2E134TMI and fluoride allows the synthesis of HPM-1 (STW) zeolites in the whole 0-1 range of Ge$_f$ molar fractions (see Table \ref{tbl:synth}). The robustness of this type of synthesis that combines the structure directing effects of both 2E134TMI and fluoride ions is revealed not only by the full Si-Ge substitutional range attainable but also by the fact that STW is the only zeolite that crystallized within a relatively wide range of crystallization times. This clearly reveals the superior structure-directing effect of 2E134TMI compared to the original organic SDA (diisopropylamine), which produced a mixture of phases,\cite{Tang2008} or a more recent SDA (N,N-diethylethylendiamine) which produced a pure STW phase in a limited range of conditions (particularly regarding the Ge$_f$ compositional range).\cite{Zhang2015a} For another recent SDAs based in the imidazolium ring (pentamethylimidazolium), only the synthesis of either pure silica STW or of intermediate germanosilicates and germanoaluminosilicates have been so far reported.\cite{Schmidt2014,Brand2017,Lu2018} We also point out that the synthesis of pure \ce{GeO2}-STW had never been described before. 

The combined structure-directing ability of 2E134TMI and \ce{F-} is likely helped up to some extend by the tendency of Ge to produce zeolites containing D4R (a structure-direction tendency shared with fluoride). However, and somehow surprisingly, the crystallization of the pure Ge-end member appears to be the less favorable one within the series, since only in that case we observed noticeable deviations from the noted crystallization of STW (see last five entries in Table S2): at short times (27 hours) we collected no solids by filtration or centrifugation, while at long times (over 100 hours) a dense quartz-like phase, and latter an argutite-like phase, started to compete. We also observed some reproducibility problems at Ge$_f$=1, since in two different runs we obtained either a very small yield of pure HPM-1 at 113 hours or HPM-1 with some quartz-like \ce{GeO2} in a higher yield at 96 and 102 h.

\begin{figure}
\centering
	\includegraphics[width=8.8cm]{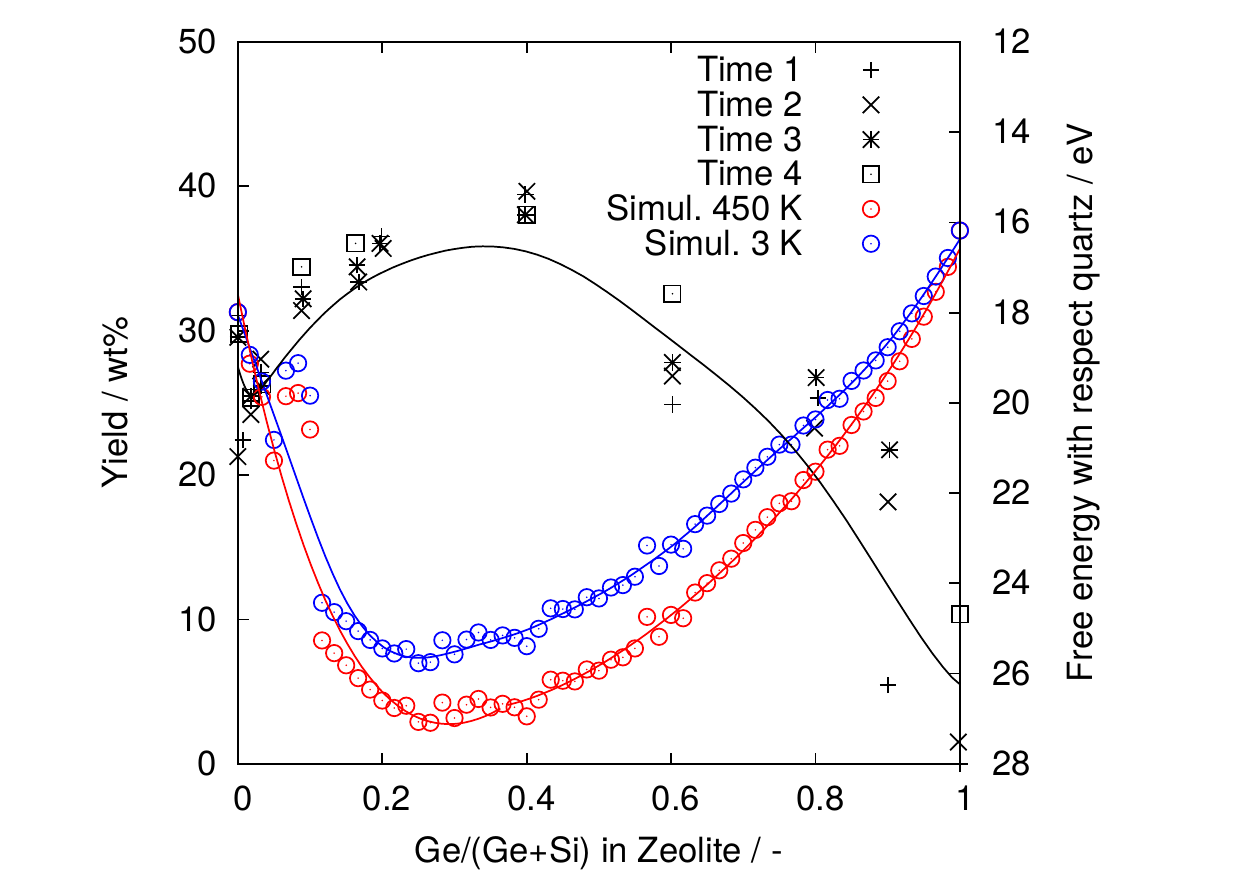}
	\caption{Yield of solids as a function of the Ge molar fractions in the zeolites (black dots). At any given Ge$_f$ different data markers refer to different crystallisation times. B\'ezier black curve fitted from experimental points is shown in the figure as a guide to the eye. Free energy as a function of Ge$_f$ at 450~K and 3~K, red and blue circles, respectively. Red and blue solid lines represents non-linear fittings of a potential function.}
    \label{fgr:yield}
\end{figure}

At the more Si-rich side of the series, STW is the only crystalline phase produced and its crystallization markedly accelerates when Ge substitutes for Si even in very small fractions (compare entries 1, 5 and 9 in Table S2). It is interesting that, for any of the crystallization times producing STW, the yield of zeolite goes through a maximum as Ge$_f$ increases and then decreases significantly, so that the value for the pure germanate material is much lower than for the pure silicate end member, Figure \ref{fgr:yield}. The oxide-based yield shows even stronger differences: the value for the Ge-end member is less than half that of the Si-end member and less than a quarter of the maximum at Ge$_f\approx 0.4$). 

We note (see Figure \ref{fgr:fit} in the Supporting Information) that the relationship between the Ge molar content in the gel and in the zeolite is almost linear, but there are small deviations that perfectly match a non-linear fit to a 3rd order polynomial. These deviations are however small in magnitude and cannot explain the observed change in yield. Hence, in order to get a deeper understanding of the factors that influence the observed volcano-type curve of the synthesis yield, we calculated the free energy of formation of the zeolite, which is plotted in Figure \ref{fgr:yield} as a function of the Si/Ge molar fraction. As shown in the figure, the yield curve has an asymmetric shape and is in close agreement with the inverse of the free energy of formation predicted by the atomistic modelling, i.e. the position at which the yield reaches the maximum value matches the minimum of the free energy curve. Thus, the observed change in yield responds to the change in free energy, as  expected. It is also worth noting that the entropic contribution does not affect in a significant way the shape of the free energy curve, since the simulation results are very similar for temperatures as different as 3 and 450 K.

\subsection{Characterization}

\begin{figure}
\centering
 \includegraphics[width=8.8cm]{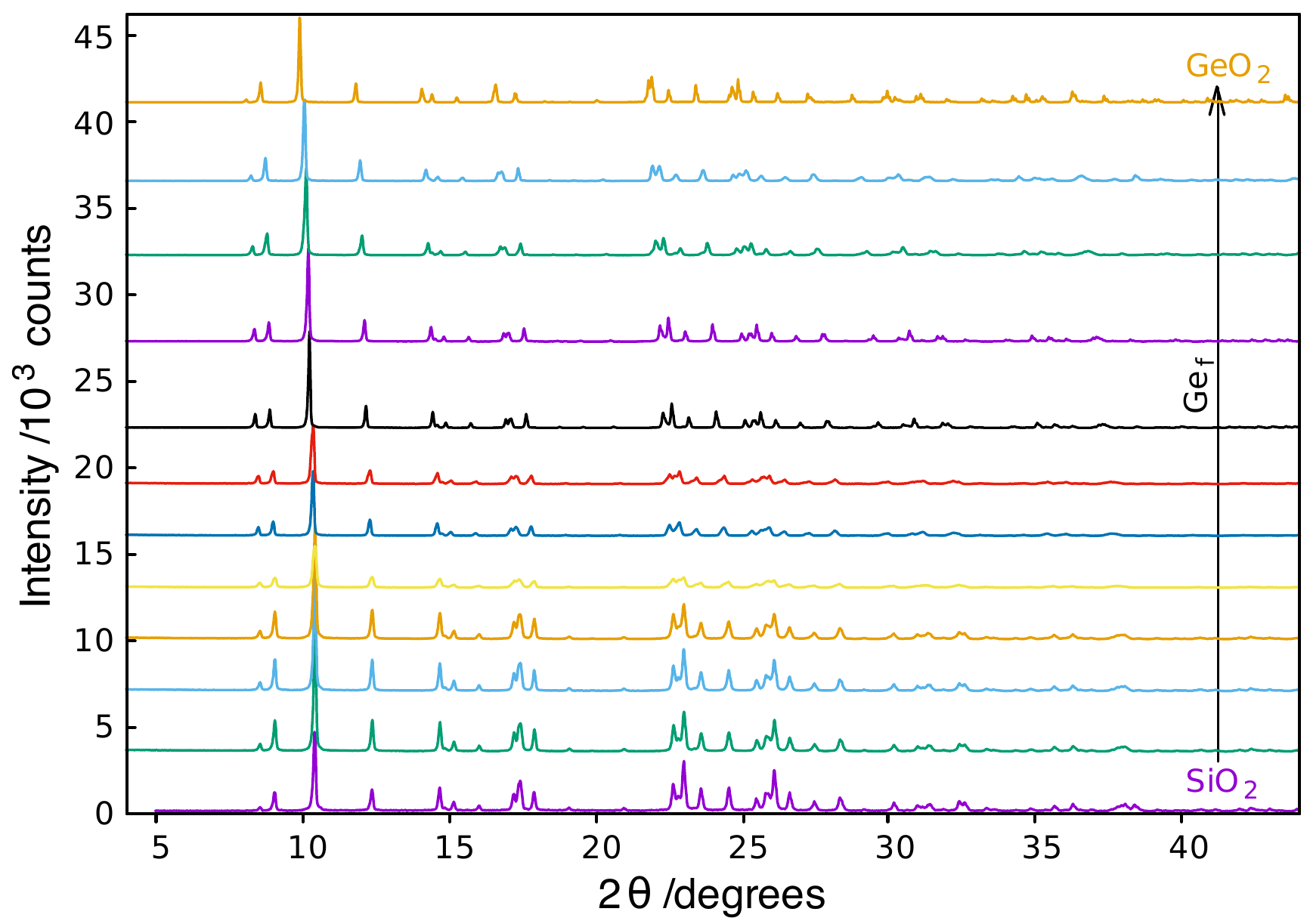}
 \caption{Powder XRD of the (Ge,Si)-HPM-1 series with varying germanium molar fractions in the gel (from bottom to top) Ge$_f=0.00, 0.009, 0.019, 0.032, 0.09, 0.167, 0.20, 0.40, 0.60, 0.80, 0.90$ and $1.00$. All the samples were crystallized at 175 $^\circ$ C for 144 hours, except the top one, pure \ce{GeO2}-HPM-1, which was crystallized for 113 h.}
 \label{fgr:xrd}	
\end{figure}

The powder XRD patterns of the STW samples, Figure \ref{fgr:xrd}, display clear changes in the positions of the different reflections as the composition changes. This is, in principle, as expected because of the Ge substitution for Si and the different size and different T-O lenghts of Si and Ge. There are abundant examples in the literature of close to linear changes of unit cell parameters as a function of T-atom substitution,\cite{Dwyer1991,Camblor1992,Sulikowski1996} although at least one exception showing a reversal of the expected trend also exists.\cite{Camblor1996f}

\begin{figure}
\centering
	\includegraphics{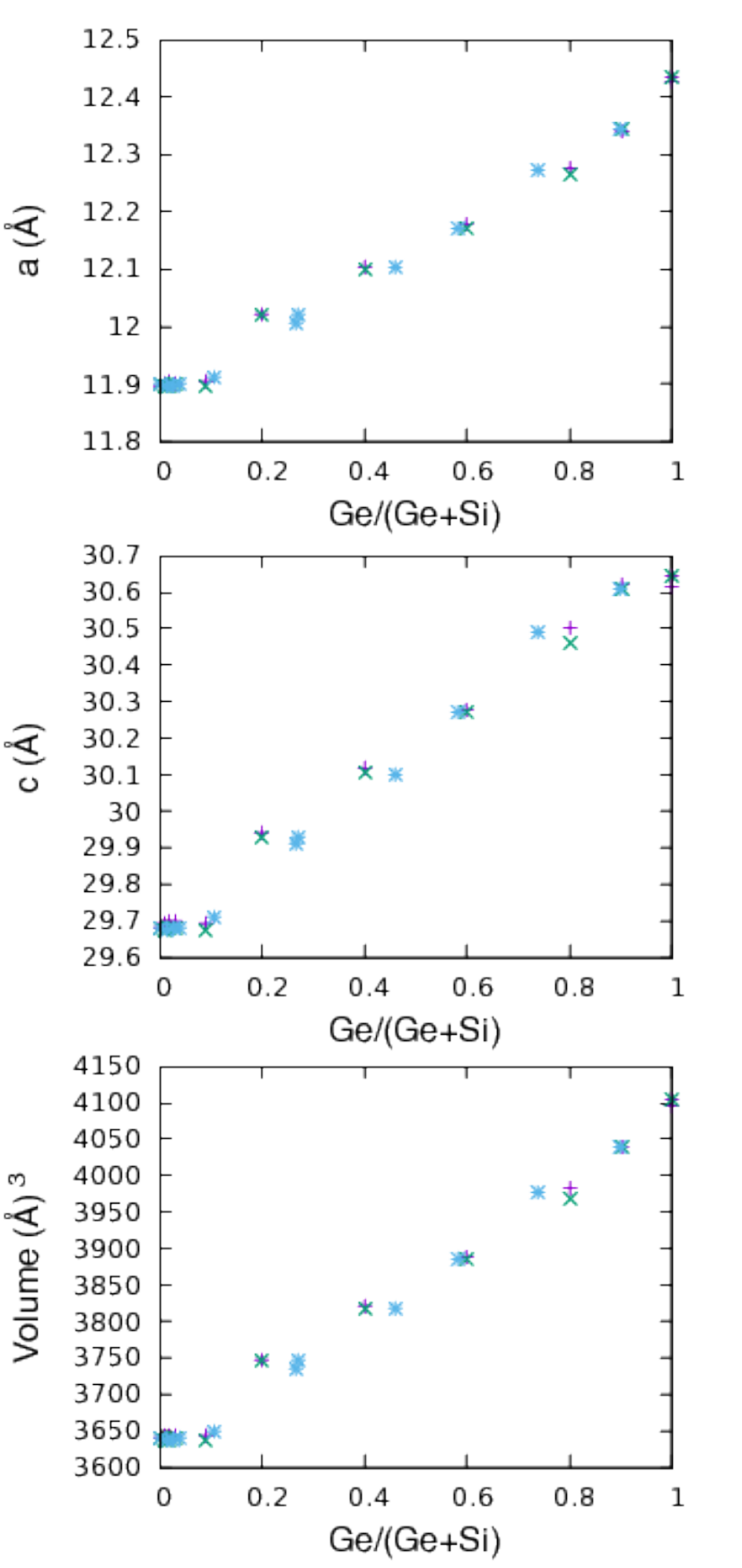}	
		\caption{Variation of the unit cell edge $a$ (top), the unit cell edge $c$ (middle), and the volume $V$ (bottom) of (Ge,Si)-STW as a function of the Ge fraction in the gel. Purple + and green x markpoints correspond to samples crystallized for 24 and 144 hours,respectively. Blue asterisks correspond to Ge$_f$ experimentally measured by ICP. }
    \label{fgr:acV}
\end{figure}

In the case of Si,Ge-STW, the overall trend is the expected expansion as the Ge fraction increases and the correlation is indeed close to linear, specially for Ge$_f>0.2$, for both the unit cell edges size and volume. However, a carefull inspection  at the high silica side of the series shows little, if any, noticeable change in the bottom five traces of Figure \ref{fgr:xrd}. In fact, the refined unit cells do not change appreciably for small substitutions of Si by Ge (Ge$_f<0.2$). As seen in Figure \ref{fgr:acV}  the overall increase in $a$, $c$ and $V$ from the pure silica to the pure germania end members is of around $4.4$, $3.3$ and $12.6$\%. However, in the Ge$_f$ range from 0 to 0.1 there are essentially no changes, instead of the expected increase in unit cell to around $a=11.95$ \AA, $b=29.78$ \AA~and $V=3690$ \AA$^3$ if the overall trend were followed.

We propose that, since the [\ce{GeO4/2}] tetrahedron is larger but also more flexible than the [\ce{SiO4/2}] tetrahedron, small amounts of Ge can enter the framework without significantly altering the unit cell size. This buffering effect appears to occur in the 0-0.1 range and contrasts with the relatively large changes in the $^{29}$Si and specially $^{19}$F spectra of the same samples (see below). This could be related to a preferential sitting of one Ge atom in each D4R in this Ge$_f$ range, see below. In contrast, in Ge-MFI, lacking D4R, \citeauthor{Kosslick1993} found a significant increase in the cell parameters in the 0-0.13 Ge$_f$ range.\cite{Kosslick1993}

\begin{figure*}
\centering
 \includegraphics[height=11cm]{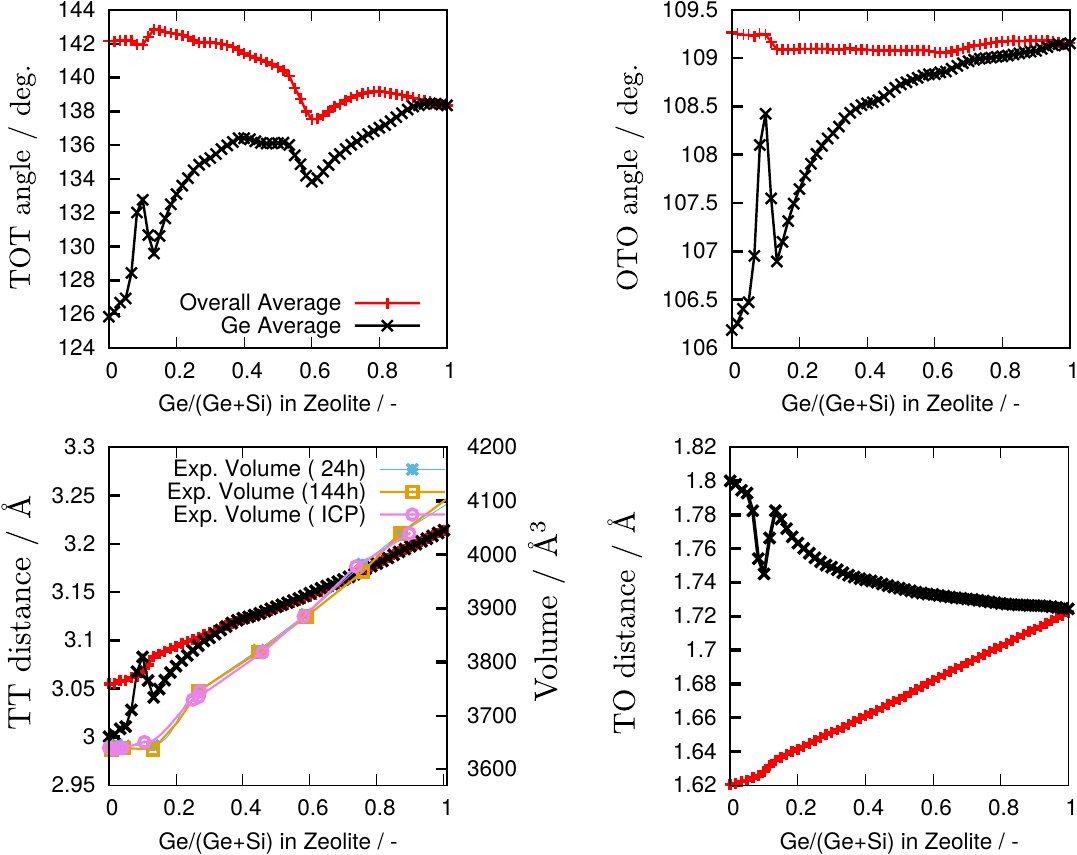}
 \caption{TT averaged distance (Bottom Left), TO average distance (Bottom Right), TOT average angle (Top Left), and OTO average angle (Top Right) versus Ge$_f$. Red and black points represent the overall average and average which involve tetrahedrons with Ge atoms, respectively. }
 \label{fgr:prop}
\end{figure*}

In order to understand the reasons for the observed changes in cell parameters and volume, we made use of lattice energy minimisation based geometry optimization, which give us an atomistic insight into the system, in direct connection with the local structure. There are some limitations to the accuracy that our simulations can provide, associated mainly to the use of force fields, and the non-inclusion of factors such as temperature and the presence of SDAs in the structure. But despite those limitations, we found a good agreement between simulation and experimental data (see Figure \ref{fgr:cell}), which gives us confidence in the atomistic behaviour of the simulated system. To extract the relevant information, we plot the changes of T-T and T-O distances, as well as T-O-T angles, as a function of the Ge content. Since the Ge$_f$ in the zeolite was close to that in the gel but not completely identical, we performed a non-linear fit of the data (see Figure \ref{fgr:fit}) and represent the experimental cell values as a function of the calculated Ge$_f$ in the zeolite. As we can see in Figure \ref{fgr:prop}, the fact that, at low Ge contents, the cell volume remains constant (as observed in Figure \ref{fgr:cell}), can be explained by the ability of a Ge tetrahedron to adapt to local deformations. In order to better understand the curves shown in Figure \ref{fgr:prop}, we remind that the original unit cell has 60 Si atoms, so that the fully Ge-substituted system is achieved after the 60 additions represented by the 60 crosses of the figures. The first cross corresponds to the substitution of just one Ge atom. It is noticeable that, at very low Ge contents, both distances (Ge-O) and angles (O-Ge-O and Ge-O-T) show singularities, which can be understood as follows. The introduction of only one Ge in each of the six D4R cubes is easily accommodated by the structure, as the effect of the enlarged T-O distance is compensated by reducing the T-O-T angles, as shown in Figure \ref{fgr:D4R_types}. It is important to note that the presence of a F atom nearby the Ge atom causes larger distortions than those expected by the sole effect of the introduction of a Ge atom. While exploring the conformational space towards the global minimum energy, we noticed the presence of multiple local minima along with the deformation of the Ge bearing tetrahedra, indicating the presence of a complex potential energy surface. The small relative depth of the energy wells served as an indication of accessible metastable phases and of the probability of transitions, which suggests dynamic flexibility behavior, in agreement with previous findings.\cite{Verheyen2014,GutirrezSevillano2016,Bueno-Perez2018} This suggests that modelling the structural features of Ge containing zeolites, even with a static view, provides means for predicting the flexibility, if any, of these materials. The addition of a second Ge atom in a D4R cube brings a large asymmetry into the local structure around the Ge atoms, since only one of the two Ge atoms is in close contact with the F atom (at ca. 1.82 \AA), while the other is further away (ca. 2.75 \AA). As a result, the former Ge atom behaves, from a local structure point of view, similarly to an isolated Ge-F pair, and the later behaves like an isolated Ge atom. We thus anticipate that the resilience to modify the cell upon a small extent of Si substitution by Ge may be a rather general behaviour for D4R-containing zeolites (we could not find relevant crystallographic data in the literature for this low Ge$_f$ range, though).

The infrared spectra of a series of as-made STW zeolites prepared from gels with different Ge$_f$ are shown in Figure \ref{fgr:ir}. Apparently, the overall effect of the presence of Ge is to cause a new set of vibrational bands at lower wavenumbers, rather than simply redshifting the bands.

Figure \ref{fgr:fesem} shows that as the Ge$_f$ increases the crystal habit changes in the sense of gradually reducing the prismatic faces. Thus, the 'double tip pencil' habit (i.e. hexagonal prisms ending in hexagonal pyramids) characteristic of pure silica and very high silica HPM-1 almost completely disappears for Ge$_f$  $\geq$ 0.2, which consists of hexagonal bipyramids. 

The thermogravimetric analyses of HPM-1 solids prepared at different Ge$_f$ ratios are provided in the supplementary information (Figure \ref{fgr:tg}). As the Ge fraction increases the weight losses decrease, as expected for the larger atomic mass of Ge compared to Si. Further, the temperature of the main weight loss also increases and, for the higher Ge$_f$ values, several weight gain events are clearly observed (starting mainly around 700 $^\circ$C and again around 900 $^\circ$C), both effects likely resulting from the complex nature of oxidation--reduction processes that Ge-containing zeolites typically undergo (notably including framework \ce{GeO2} reduction and reoxidation) as very recently reported.\cite{Villaescusa2016a}

\subsubsection{Multinuclear NMR}
\begin{figure}[h!]
\centering
	\includegraphics[width=8.8cm]{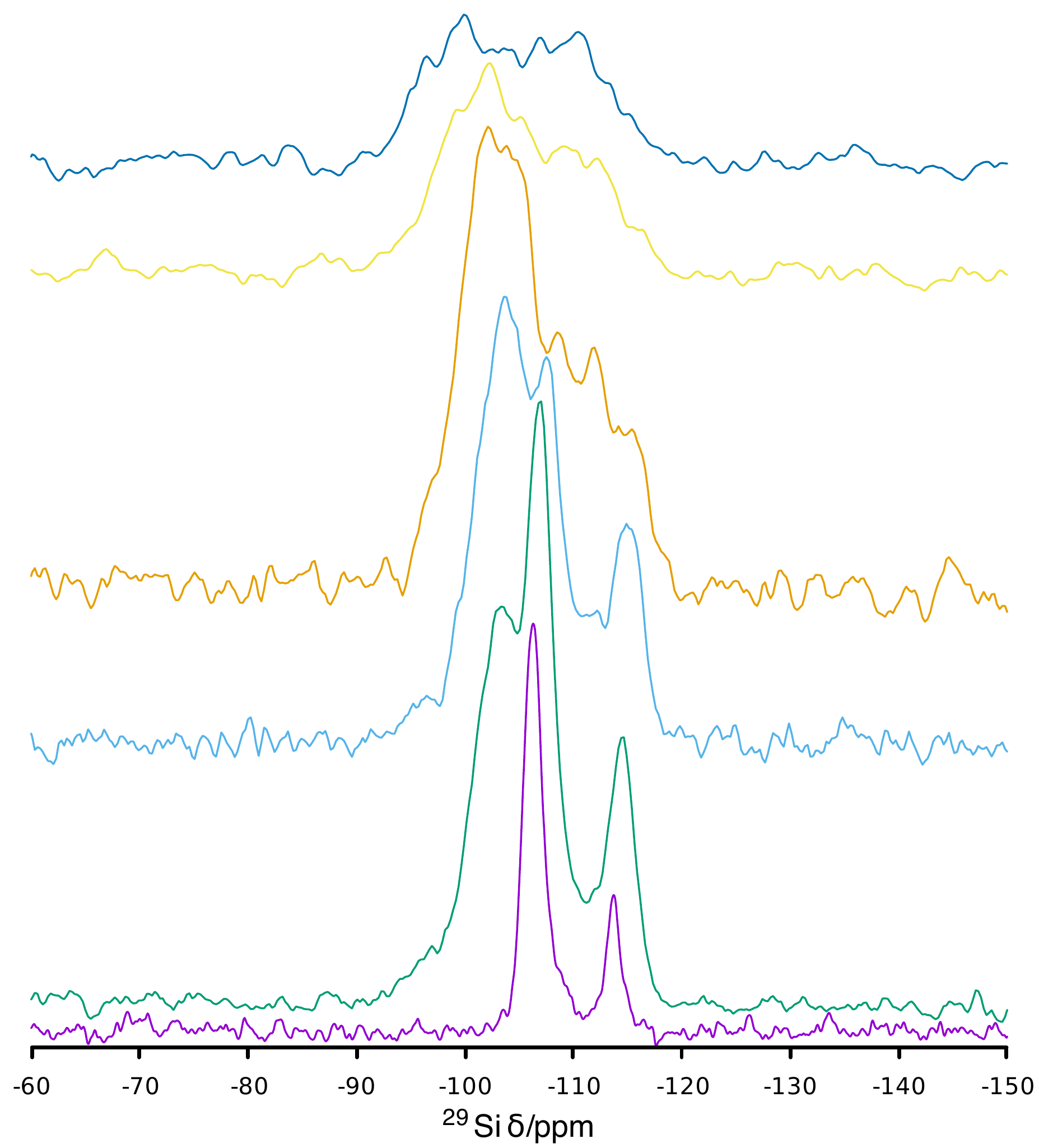}
		\caption{$^{29}$Si MAS spectra of (Ge,Si)-STW with Ge$_f$= 0.00, 0.166, 0.20, 0.40, 0.60 and 0.80(from bottom to top). The lower trace is a direct irradiation spectrum while the rest were collected under cross polarization.}
    \label{fgr:29sicp}
\end{figure}

$^{13}$C and $^1$H MAS NMR spectra (not shown) demonstrate the organic SDA is occluded intact in the zeolites. The $^{29}$Si CP MAS NMR spectra of several relevant germanosilicate HPM-1 samples are shown in Figure \ref{fgr:29sicp}. The lower trace in the figure is the direct irradiation $^{29}$Si MAS NMR spectrum of the pure silica material, which shows two clear resonances at -106.2 and -113.6 ppm, with a relative intensity ratio close to 4:1, assigned to Si in crystallographic sites T$_{1-4}$ and T$_5$, respectively.\cite{Rojas2013b} These correspond to sites in and out of D4R, respectively. Interestingly, introduction of Ge causes a new resonance to appear at lower fields (ca. -103.2~ppm for a Ge$_f$=0.167). If Ge shows a preference to occupy sites within D4R units, see below, the new resonance at -103.2 ppm may be adscribed to \underline{Si}\ce{(OSi)3OGe} sites in D4R. \citeauthor{Schmidt2014b} and \citeauthor{Whittleton2018} found a similar downfield shift for Si(Si$_3$,Ge) (-102 ppm and -104~ppm, respectively) compared to Si(Si$_4$) resonances (-108 ppm) in silicogermanate LTA zeolite.\cite{Schmidt2014b} On a close inspection, that resonance shows significant intensity almost down to -90 ppm, suggesting the existence of overlapped \underline{Si}\ce{(OSi)2(OGe)2} resonances. 
For Ge$_f$=0.20 all the resonances experience a small upfield shift (to 103.7, -107.5 and -114.9~ppm, respectively). Upon a further increase in the level of Ge for Si substitution to Ge$_f$=0.40, the lower field signal is the dominant one and clearly consists of several resonances, while the high field side of the spectrum consists of at least three heavily overlapped resonances (-108.4, -111.8 and -114.9~ppm, respectively, suggesting site T5, not belonging to D4R, may be now populated by \underline{Si}\ce{(OSi)2(OGe)2}, \underline{Si}\ce{(OSi)3OGe} and \underline{Si}\ce{(OSi)4}, respectively. This is not unexpected if the fraction of Ge in D4R sites is significant. As the Si content decreases further, the spectra becomes much broader, blurry and noisy. We cannot perform a more quantitative analysis of the spectra because the intensities in the CP spectra depend on the proximity to protons and the direct irradiation $^{29}$Si MAS NMR spectra require prohibitively long recycle delays to achieve spectra with decent signal to noise ratios (see Figure \ref{fgr:29sidirecto}).

\begin{figure}[h!]
\centering
	\includegraphics[width=8.8cm]{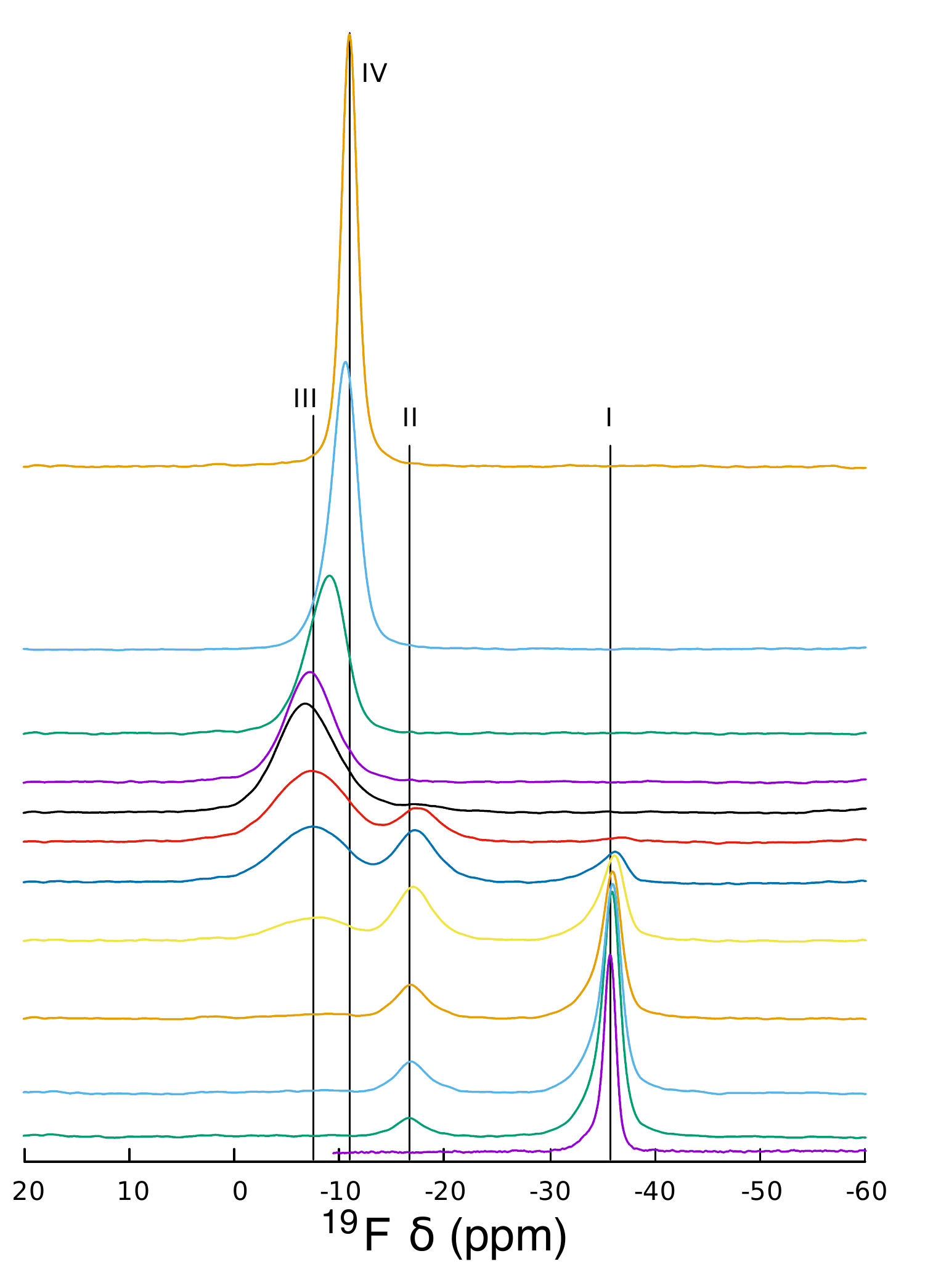}
		\caption{$^{19}$F MAS NMR spectra of (Ge,Si)-STW with Ge$_f$= 0.00, 0.009, 0.019, 0.032, 0.09, 0.166, 0.20, 0.40, 0.60 and 0.80, 0.9 and 1.00(from bottom to top). The four types of resonances found are marked with the vertical lines I-IV, placed at the position where they appear at lower Ge$_f$ (except line IV, placed at the resonance of the pure germanate.)}
    \label{fgr:19f}
\end{figure}

More interesting for a better understanding of these materials is the $^{19}$F MAS NMR spectroscopy (Figure \ref{fgr:19f}), which is much sensitive to the type of cavity in which fluoride resides and to its kind of interaction with framework atoms. In the case of zeolites containing D4R, fluoride is typically occluded within this small cavity, and its chemical shift depends on the composition of the D4R. The pure silica HPM-1 material displays a single resonance at around -35.7 ppm, that we will call here resonance 'I'. The assignment of resonance I to \ce{F-} occluded in purely siliceous D4R (i.e., 8Si,0Ge D4R) is well stablished for several pure silica zeolites,\cite{Caullet1991, Villaescusa1999,Camblor2001,Barrett2003} as well as for pure silica STW,\cite{Rojas2013b} despite the fact that in this zeolite it appears quite downfield shifted compared to more typical values (-37/-40ppm).\cite{Camblor2001} When Ge is introduced in the synthesis mixture, even in minute amounts (Ge$_f$= 0.009), a new resonance appears around -16.6ppm (resonance 'II'). As more Ge is introduced, this resonance first increases, then decreases in intensity, while it experiences an upfield shift (up to -17.5 ppm). Upon increasing the Ge content above Ge$_f$ = 0.032 a broad resonance ('III') appears around -7.5 ppm and increases in intensity while shifts to lower field up to Ge$_f$=0.4. We note here that the very significant changes just commented occur in a range of Ge$_f$ in which the zeolite framework appears to be reluctant to expand, as mentioned above (see Figure \ref{fgr:acV}). Then, an apparent upfield shift starts, while the resonance becomes narrower. This apparent change in shift, the narrowing of the band, and prior literature reports on other zeolites lead us think that there is a fourth resonance (IV, around -10/-11 ppm), rather than one that first moves downfield then jumps upfield, and that at intermediate Ge fractions resonances III and IV severely overlap. Since the pure germanate end-member displays a single, relatively narrow and pretty symmetrical resonance IV at -11.0 ppm, we would initially assign it to \ce{F-} occluded in D4R built only of Ge and O (i.e., 0Si,8Ge D4R) (literature values vary roughly in the -9 to -16 ppm).\cite{Villaescusa2002a}
\begin{figure*}[htp!]
\centering
    \includegraphics[width=18.3cm]{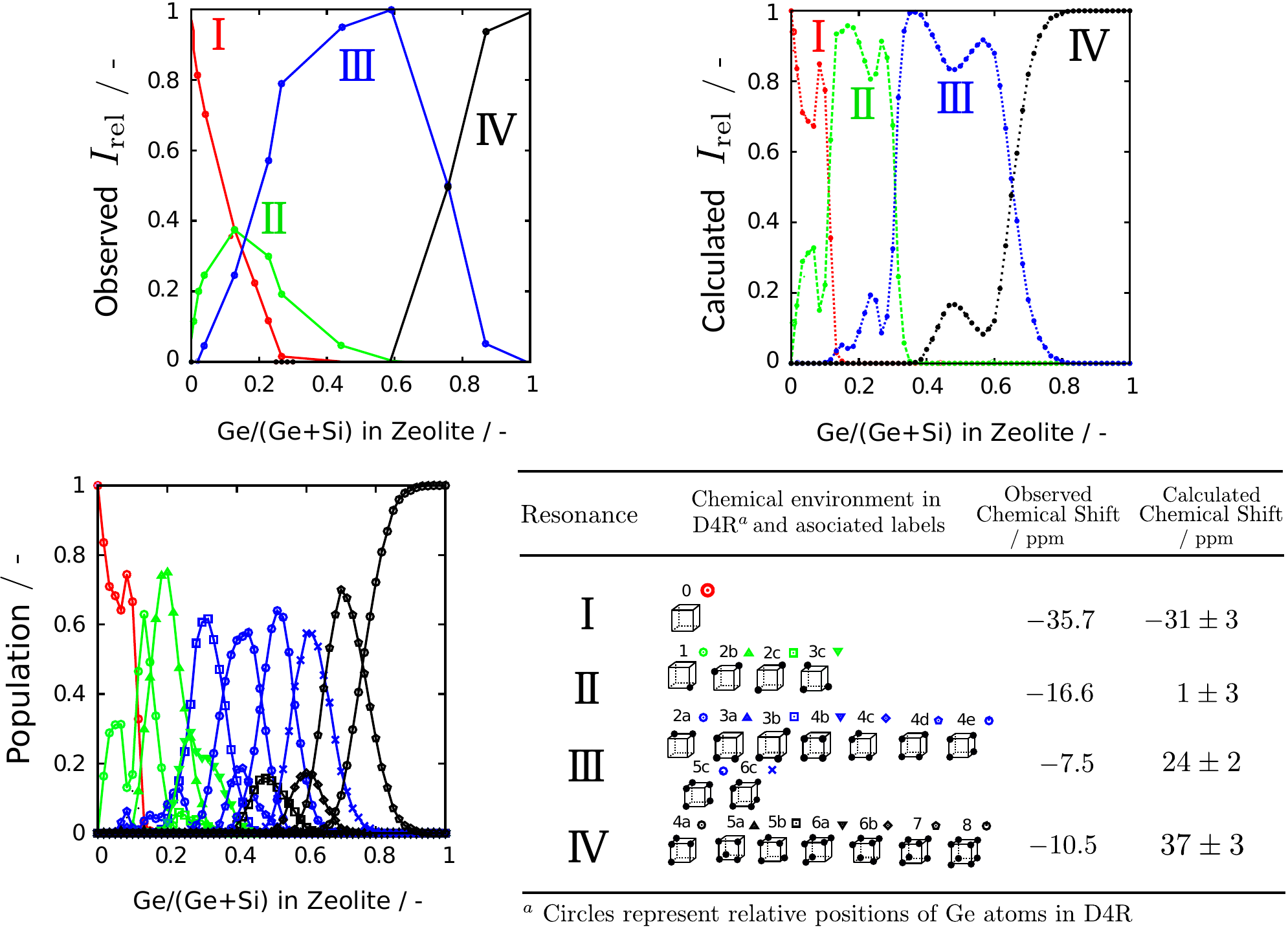}
	\caption{Top: Relative intensities of the four resonances observed in the $^{19}$F MAS NMR spectra of (Ge,Si)-STW zeolites as a function of the Ge fraction in the zeolites, Ge$_f$ (left: experimental; right: convolution of calculated populations of chemical environments in D4R, grouped by similar calculated chemical shift). Bottom-Left: Calculated population of chemical environments in D4R as a function of Ge$_f$. Bottom-Right: Our proposed assignment of observed $^{19}$F MAS NMR resonances to chemical environments in D4R, with experimental and averaged calculated chemical shifts.}
    \label{fgr:theore_D4R_distribution}
\end{figure*}

The assignment of the remaining $^{19}$F resonances, II and III, is intriguing and has been the subject of debate. There are typically four types of resonances in the  $^{19}$F MAS NMR spectra of (Si,Ge)-zeolites containing D4R, despite the fact that, in principle, there may be up to nine different Ge contents in a D4R unit (from 0 to 8) and for several of these contents there may be a number of different configurations of Si and Ge within the D4R, see below. The scarce number of resonances could be just due to resonance overlapping, to some configurations being prohibited or scarce or to an insensitiveness of $^{19}$F to certain differences among configurations and compositions. \citeauthor{Sastre2002a} assigned resonances at -38, -20 and -8ppm in silicogermanates ITQ-17 and ITQ-7  to \ce{F-} in nD4R with, respectively, 8, 7, and either 5 or 6 Si atoms, being more favorable to 5.\cite{Sastre2002a,Blasco2002} \citeauthor{Wang2003} studied octadecasil silicogermanates synthesized with three different SDA cations, covering for two of them the whole range of Ge$_f$ from 0 to 1. The pure silica and pure germania end members present resonances at around -38 and -15 ppm, which are thus assigned to \ce{F-} in D4R with 8 and 0 Si, respectively. For intermediate compositions resonances around  -8 and -19 ppm were assigned to the presence of 4 and 6 Si per D4R, respectively, and the authors concluded there is an ordered pattern of Ge insertion in the D4R units in which Ge-Ge pairing tend to be avoided.\cite{Wang2003a} To complicate things, each one of these resonances may change position depending on the Ge$_f$,\cite{Wang2003} or ocluded SDA cation.\cite{Villaescusa2016}

Latter on, \citeauthor{Sastre2005} suggested that there may exist direct covalent Ge-F bonds in D4R units, with expansion of the coordination of the involved Ge to 5.\cite{Sastre2005} The same authors calculated the chemical shifts of fluoride occluded in different configurations of D4R units containing 0, 1, 2 , 3, 4 or 8 Ge atoms (but they did not consider 5, 6 or 7, for undisclosed reasons) and concluded that, due to the displacement of fluoride out of the cage center and towards a corner, the main factor determining the chemical shift of fluoride was the nature of the 4 closest T, i.e., the n number of closest Si and m number of closest Ge, with $n+m=4$.\cite{Pulido2006} Thus, the chemical shift of fluoride increased (values more positive) when the number m of Ge atoms closer to F increased. This could explain that fluoride in D4R containing 4Si4Ge would resonate at a similar chemical shift as those in 5Si3Ge. This view differs significantly from that of \citeauthor{Wang2003a} described above,\cite{Wang2003a} because if Ge-Ge pairings were avoided the 4Si4Ge D4R unit would have $m = 1$ Ge as closest neighbours to F.

After deconvolution of the assumed III+IV resonance in the spectra of materials with Ge$_f$ in the 0.6--1.0 range, the evolution of the four resonances as a function of the Ge content in the gel is shown in Figure \ref{fgr:theore_D4R_distribution}, top left, solid lines. It is worth noting that, at the high silica side of the substitutional series, the $^{19}$F resonances change very drastically as the Ge$_f$ increases. The sharp decrease of resonance I, which is replaced by resonance II and then III occurs in a range of Ge$_f$ that, as discussed above, shows essentially no variation in unit cell parameters. For a Ge$_f$ of 0.09 the spectrum consists of resonances I, II and III with roughly similar intensities ($\approx$ 37:38:25), while the unit cell shows essentially no variation in dimensions. 

At first sight, there would be little question about the assignments of resonances I, II and IV. In the case of resonances I and IV the assignment would apparently get strong support from the spectra of the pure \ce{SiO2} and \ce{GeO2} end members, respectively, in which resonances I and IV have no other possible assignment. In the case of II, its appearance at very low Ge$_f$ together with its fast grow and decay as Ge$_f$ increases, would apparently support the assignment to 7Si1Ge-D4R. Resonance III is, obviously, a problem, and the fact that there is only one such resonance implies that either there is a very ordered pattern of Ge introduction (as proposed by \citeauthor{Wang2003a})\cite{Wang2003a} or at least this resonance actually consists of several resonances overlapped (as suggested by \citeauthor{Pulido2006}).\cite{Pulido2006}

We made use of DFT calculations in an attempt to shed light on the origin of the four different resonances. It turned out (Figure \ref{fgr:theore_D4R_distribution}) that the assignments are likely much more complicated than as presumed above or as described in the prior literature. The computed chemical shifts cluster around four well-defined ranges, which can be reasonably well matched to the observed $^{19}$F resonances.
We have label the four set of resonances that result from the calculations using the same numerals of the analogous experimental resonances, i.e. accordingly to the order the resonances appear in the experiments as the Ge content increases.
It is interesting to note that, except resonance I (associated to pure silica units), each resonance has contributions from various configurations, including some with different distributions of Si-Ge atoms in the D4R units (Figure \ref{fgr:theore_D4R_distribution}, bottom-right). Based on the dispersion of the computed values of the chemical shifts, we assign the observed four resonances to F atoms occluded in D4R cages with the following Ge content and Ge clustering pattern:
\begin{enumerate}
\item \underline {no Ge} atoms (resonance I)
\item \underline {isolated Ge} atoms, i.e. Ge atoms with three Si atoms as cage neighbours (resonance II). Thus, there may be 1-3 Ge atoms in the D4R.
\item \underline {Ge pairs} not satisfying the conditions of type IV (resonance III). Here, up to nine different configurations with 2-6 Ge atoms are possible. 
\item \underline {closed Ge clusters}, i.t. configurations with at least one Ge possessing three Ge atoms as next nearest neighbours (resonance IV). There may be seven different such configurations with 4-8 Ge atoms.
\end{enumerate}

The proposed assignment is not simply based on the number of Ge atoms in the D4R unit, but mainly on how Ge atoms are distributed in that unit with regard to F, so that the very local chemical environment of the F atoms is constant within each group responsible for the four resonances. For example, the F atoms in D4R configurations \emph{1} and \emph{2b} (Figure \ref{fgr:theore_D4R_distribution}, bottom left) have similar chemical environments, since the core electrons of the F atom in configuration \emph{2b} are being affected mainly by just one Ge atom, as if it were in a 1 Ge-containing D4R. Accordingly, in configuration \emph{2a}, in which the two Ge atoms are nearest neighbors, the core electrons of the F atom sense the simultaneous presence of two Ge atoms. As we have noted in the methodology, the covalency of the F environments in the D4R is expected to affect the NMR calculation. The deviation of the predictions with respect to the experimental values increase with the number of Ge presents in the nearest environment of fluoride atoms in the D4R, i.e. the covalency, maintaining the statistical error of the calculations constant.  The deviation is proportional to the numeral of the resonance, and it is about a constant value of 15 ppm (see Figure \ref{fgr:localenvironments}).
Note that the apparent uncorrelated calculation of the chemical shift of such large system using the PBE functional acquire rationalization through this analysis and provides a means to understand and predict chemical shifts of F in those structures.

It is worth noting that the presence of resonance III at the high silica side of the substitutional series (very dimly at Ge$_f$=0.032 but robustly evident at Ge$_f$=0.09) implies significant formation of Ge-O-Ge bonds at relatively low Ge contents, according to our assignment. We find this, however, rather logical because, as we will see below, sites T1 and T2 are clearly the preferred sites to be occupied by Ge and they are multiply connected to each other (each T1 or T2 connects to one T1 and one T2). Then, Ge-O-Ge formation at low Ge$_f$ suggests that there must be no significant penalty for Ge pairing in zeolites, as opposed to prior suggestions of a Ge-Ge avoidance at Ge$_f$\textless 0.5.\cite{Wang2003a}

Supported by the good agreement observed between the synthesis yield and the computed free energy of the solids, we use these energy values to construct the set of representative Ge configurations for each Ge content, covering the whole Ge-containing  compositional range, i.e. from 1 to 60 Ge atoms per unit cell. We considered the configurations whose sum of occurrences of probability are at least 99.9 \%. Taking into account the assignation of the DFT-computed NMR resonances, as well as the chemical environments identified, we recreated the theoretical population of the F-NMR resonances. The comparison between theoretical resonances and experimental intensities of the observed resonances is qualitatively good, as can be observed in Figure \ref{fgr:theore_D4R_distribution}, top left, dotted lines, which suggests that the distribution of chemical environments is reasonably well represented in the space of possible configurations that the effective Hamiltonian predicts (Figure \ref{fgr:theore_D4R_distribution}, top right).
In this respect, differences between theoretical and experimental findings may be due to heterogeneity in the composition of the D4R.
For example, for Ge$_f=0.5$ a small contribution for resonance IV appears in the calculated population curve (Figure \ref{fgr:theore_D4R_distribution}-top right) that is absent in the experimental spectrum. Note that this calculated intensity for the resonance IV is again retrieved for Ge$_f=0.65$, which supports the view of heterogeneity in the composition as the main source of the observed small differences. Inaccuracies of the used interatomic potential, not being fine enough to see the small variations in Si-Ge distribution within the D4R units might be also a source for the mismatch.
\begin{figure}[h!]
\centering
	\includegraphics[width=8.8cm]{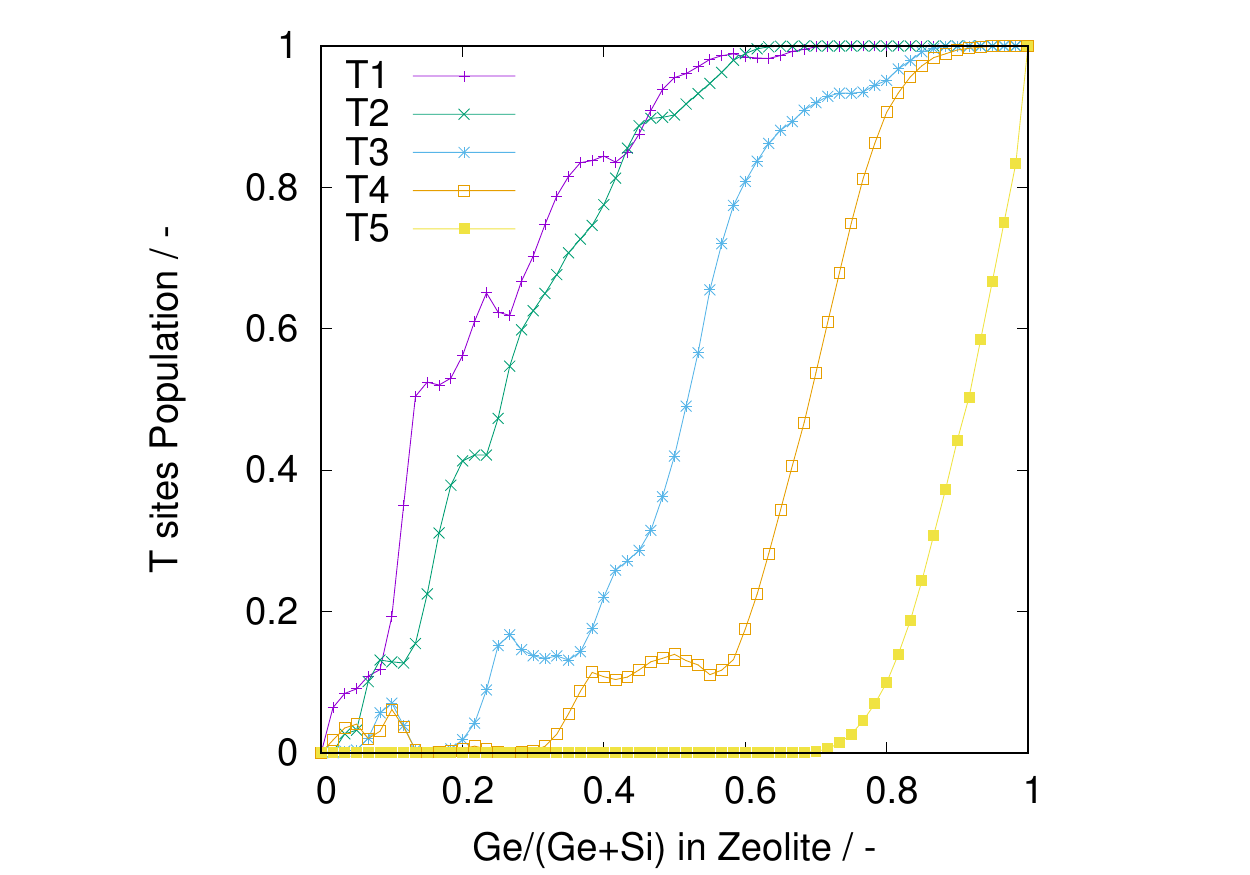}
    \caption{Variation of the populations of the different T sites, as a function of Ge content in zeolite, Ge$_f$, calculated with the EH.}
    \label{fgr:pop}
\end{figure}

We can also get an insight into the different preference of Ge (Si) to occupy the different crystallographic T-sites. Employing an effective Hamiltonian, we obtained the population of each T site as a function of the Ge$_f$, as shown in Figure \ref{fgr:pop}. According to these calculations, the Ge preferential occupation of sites goes in the order T1 $\geq$  T2 \textgreater T3 \textgreater T4 \textgreater T5. Thus, as the Ge content increases, sites T1 and T2 are occupied first. Then site T3 followed by T4 start to be occupied at relatively low Ge contents but with a lower preference over T1 and T2.Finally, after sites T1 and T2 are fully occupied but before the rest of sites in the D4R units are occupied, site T5 begins to be occupied by Ge atoms.

In the following section we shall compare the preferential occupation of the different sites predicted by molecular simulations with that obtained from Rietveld analysis.

\subsubsection{Rietveld refinement}
In order to get a deeper experimental insight into the Ge-Si substitution in HPM-1, we performed Rietveld refinements of the structures of samples prepared at Ge$_f$= 0.4, 0.6 and 1.0 using powder data collected usign synchrotron radiation, and the details are given in the Supporting Information. 
\begin{figure}[h!]
\centering
  \includegraphics[width=0.33\textwidth]{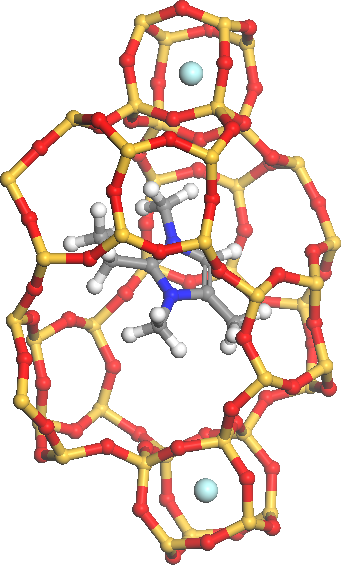}
  \caption{The organic SDA cation inside the large $[4^65^88^210^2]$ cage (only one of the two symmetrical cations is shown) and the fluoride anion inside the small $[4^6]$ cage of Ge-HPM-1.}
  \label{fgr:rietsda1}
\end{figure}

The final refined structures have reasonable unrestrained bond distances and angles (see Table \ref{tbl:riet}) and all show the fluoride anions slightly off the center of the D4R cages and closer to T1 than to any other tetrahedral atom in the framework. Figure \ref{fgr:rietsda1} shows the organic cation and fluoride anions occluded in their respective cages in the purely \ce{GeO2} HPM-1. The refined occupancies of Ge and Si in the Ge$_f=0.4$ and $0.6$ samples are close to the nominal values (0.40 and 0.57, respectively) and both show a distinct preference for Si rather than Ge to occupy T5 (the non-D4R site) and a Ge preference to occupy sites T1 and T2 over T3 and T4 (see Table \ref{tbl:occupy}). The site occupancies observed in these samples roughly agree with the order of preferential occupations determined from our calculations (see Figure \ref{fgr:pop}). In both cases, as the amount of Ge increases, T1 and T2 are populated before T3 and T4, which in turn get occupied preferentially over T5. There are, however, quantitative discrepancies between both results. The first one is that the differences in occupation between sites T1 and T2 and between sites T3 and T4  are larger in our calculations than in the experiments. And the second discrepancy is that experimentally T5 starts being slightly populated at Ge$_f=0.4$, and at Ge$_f=0.6$ its population is already roughly one third of either one of T3 and T4, while at that point T1 and T2 are not fully occupied by Ge yet. This is in clear contrast with the predicted values, which show that T5 does not start being populated by Ge until all other sites are filled, or almost filled. Roughly speaking, our calculations predict more strict preferential occupations and, hence, more abrupt changes in population trends. There are mainly two possible explanations for these discrepancies. First, we could ascribe the differences to inaccuracies of the energy calculation employed, which might be making T5 sites too unstable for Ge compared with the other sites. While this might be the case, it is also possible that we are neglecting some factors in our calculations. Namely, our calculations are based on the analysis of the thermodynamic properties of the system in equilibrium. But it is well known that kinetic factors play a relevant role in the formation of zeolites, i.e. the system might have a mixture of metastable configurations, and configurations that appear purely for kinetic reasons, and we are missing all of those in our analysis. These kinetic factors would smooth the tendencies observed in the calculations. But apart from these discrepancies, both sets of data provide the same general view, consisting in the similar Ge population of sites T1 and T2, followed by T3 and T4, and finally T5.

\begin{table}[h]  
   \caption{\ Refined Ge occupancies of T sites in Ge,Si-HPM-1}
  \label{tbl:occupy}
  \begin{tabular*}{0.48\textwidth}{@{\extracolsep{\fill}}ccc}
    \hline
    &  Overall $\text{Ge}_f=0.4$ & Overall Ge$_f=0.6$\\
    \hline  
    Ge in T1 & $0.552$ & $0.748$ \\
    Ge in T2 & $0.555$ & $0.735$ \\
    Ge in T3 & $0.388$ & $0.584$ \\
    Ge in T4 & $0.393$ & $0.591$ \\
    Ge in T5 & $0.092$ & $0.215$ \\
    \hline
  \end{tabular*}
\end{table}

\section{Conclusions}

Isomorphous substitution of Si by Ge in the synthesis of zeolite STW using 2-ethyl-1,3,4-trimethylimidazolium and fluoride affords the crystallization of the whole substitutional series from the pure \ce{SiO2} to the pure \ce{GeO2} end-members. A combined experimental-theoretical approach allowed us to get significant insight into the system, which may be of general interest for germanosilicate zeolites. As the Ge molar fraction  increases, the yield of zeolite goes through a maximum and then severely drops at the \ce{GeO2} end member. Our calculation of the corresponding free energies matches well the inverse of the yield curve. 

The isomorphous substitution of Si by Ge brings about an expansion of the structure that is roughly linear for most of the series. However, for low Ge$_f$ ($\leq$ 0.1) there is no expansion of the unit cell. This resilience to expansion is attributed to the local deformability around Ge atoms and the higher rigidity of \ce{SiO2}. 

Similarly to previously published germanosilicate zeolites containing double 4-ring units (D4R), we observe up to four distinct resonances in the $^{19}$F MAS NMR spectra, depending on the Ge content. However, the assignment of these resonances is far more complicated than previously thought. Density functional theory calculations of the $^{19}$F chemical shifts of fluoride occluded in every possible configuration of every [Si$_{(8-n)}$Ge$_n$] D4R unit (with 0 $\leq$ \textit{n} $\geq$ 8) reveals the resonances are not simply dependent on the number \textit{n} of Ge atoms but also on the extension of Ge pairing. Thus, resonances are assigned to fluoride occluded in D4R with no Ge, with isolated Ge, with Ge pairs or with Ge in closed clusters. 

Our modelling of these materials showed the presence of a complex energy surface with multiple shallow minima. We suggest that even static modelling of materials may thus provide means for predicting their flexibility.

Finally, we studied the preferential occupation of crystallographic sites by Ge both theoretically (for the whole series) and experimentally (by Rietveld refinement of structures with different Ge$_f$ using synchrotron powder diffraction data). We found a good overall agreement but with a somewhat more abrupt and sharply distinct preferences in the models than in the experimental results. This is attributed to both the limitations of the theoretical approach and to kinetic factors allowing the real existence of metastable configurations not considered by the models.

\section{Conflicts of interest}
There are no conflicts to declare.

\section*{Acknowledgements}
The authors thank the Spanish Ministry of Science and Competitiveness for funding (Projects MAT2015-71117-R and CTP2016-80206-P). R.T. Rigo thanks CAPES (Brazil) for a PhD fellowship (process 99999.012012/2013-00). S.R.G. Balestra thanks the Spanish State Secretariat for Research, Development and Innovation for his predoctoral fellowship (BES-2014-067825 from CTQ2013-48396-P). We are also indebted to the ERSF (Grenoble) and the BM25 Spline staff, particularly to G. Castro and A. Serrano as well as to L. A. Villaescusa (Valencia) for help in collecting the synchrotron XRD data and for helpful comments and suggestions. We also thank A. Valera for technical expertise (FESEM). Computing facilities by Alhambra supercomputing center (Universidad de Granada) is grateful acknowledge.

\newpage
\clearpage

\onecolumngrid
\appendix

\section{Supplementary Information available}
Ge$_f$ in zeolites as a function of Ge$_f$ in gel, Calculation of the energy of the configurations, synthesis results, cell parameters vs Ge$_f$ computed with Effective Hamiltonians, some of the energy-minimised D4Rs configurations, infrared spectra, FE-SEM images, thermograms, direct irradiation $^{29}$Si MAS NMR,difference between calculated and experimental $^{19}$F chemical shifts, Rietveld details, Rietveld plots, Table of crystallographic and experimental parameters, average bond distances and angles (Tables S1-S4 and  Figures S1-S11).

\setcounter{table}{0}\renewcommand{\thetable}{S\arabic{table}}
\setcounter{figure}{0}\renewcommand{\thefigure}{S\arabic{figure}}
\setcounter{equation}{0}\renewcommand{\theequation}{S\arabic{equation}}

\subsection{Methodology}

\begin{figure}[h]
 \centering
 \includegraphics[width=0.8\textwidth]{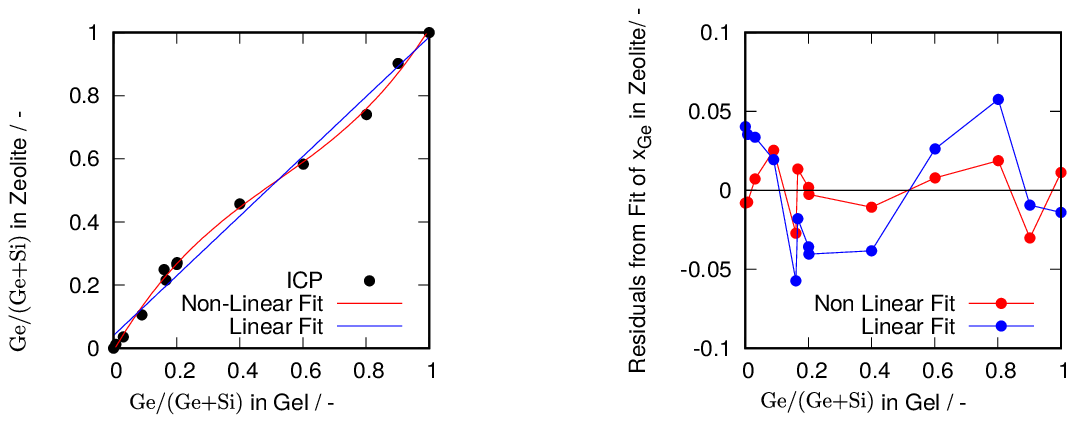}
 \caption{Molar fraction of Ge in the zeolites as a function of that in the gel.}
 \label{fgr:fit}	
\end{figure}

\begin{table}[h]
  \caption{Details of the number of configurations used for each Ge content}
  \centering
  \label{tbl:number_of_conf}
  \begin{tabular}{ccc}
    \hline
Ge/Si  & Total number  & Number of  \\ 
substitutions & of configurations & inequivalent configurations \\
\hline 
1 & 60 & \textbf{5} \\ 
2 & 1770 & \textbf{165} \\ 
3 & 34220 & \textbf{2855} \\ 
4 & 487635 & 40890/\textbf{1507}$^\emph{a}$ \\ 
    \hline
  \end{tabular}
  
$^\emph{a}$ The number of configurations considered in the calculations is shown in bold type.
\end{table}

\subsubsection{Calculation of the energy of the configurations: Effective Hamiltonian approach}

The incorporation of heteroatoms in a zeolite framework might generate a large configurational space of possible atomic distribution. For simplicity, we will concern only on binary composition, for instance in the case of the present case Si-Ge distribution. Symmetry consideration, by using the SOD program,\cite{GrauCrespo2007} allows us to map all non-equivalent configurations up to 4 Ge atoms by unit cell. Since the cell contains 60 tetrahedral sites and the symmetry of the pure silica STW zeolite framework is relatively low (space group \# 178), with 4 Ge atoms by unit cell there are already more than 40 thousand configurations. Even by using interatomic potential based calculations, this is already a heavy computational effort. For larger increase of the minority element in the binary solid solution, the number of configurations increases exponentially and therefore it is not possible to compute their energy. To deal with this, we turn to the recently developed Effective Hamilton  ian (EH) approach \cite{Arce2018}, which parametrise the atom--atom interaction in a simple numerical function. In this way, the energy of millions of configurations can be  evaluated at a small computational cost. Nevertheless, the method implies an initial high cost, since all configurations having 2 Ge (for STW a total of 165) and 3 Ge (2855 configurations) are needed first to submit to full energy relaxation, including atomic coordinates and cell parameters. Since Ge atoms confers large structural flexibility to the framework, those configurations having 4 Ge atoms, with either 4 nearest Ge neighbours or 3 Ge neighbours plus a 4th Ge atom as second next nearest neighbour, were also considered for the parameterisation of the EH. The set of zeolites considered with 4 Ge atoms has 1507 configurations.

The EH is based on consideration that the entrance of a heteroatom can be treated as a defect. First the substitution energies for isolated Ge atoms in the five distinct T sites are computed and after that, the interaction energies with the addition of new Ge atoms are computed. We therefore parameterise the effective Hamiltonian as follows:
\begin{enumerate}
\item Firstly, the perturbation energy to substitute a Ge atom, $\Delta E(\vec{r}_i)$, is calculated using the Mott-Littleton methodology \cite{MottLittentonCatlow}, for each unique tetrahedral site $\vec r_i$.
\begin{equation}
 \Delta E(\vec{r}_i)=E_i-E_0
\end{equation} where $E_0$ and $E_i$ are the lattice energies computed with GULP using the interatomic potential of pure silica structures and structures with one Ge/Si substitutions.
\item We then consider a pair interaction energy denoted as $\Delta E(\vec{r}_i,\vec{r}_j)$, where $\vec r_i$ and $\vec r_j$ are two tetrahedral sites, and is computed as the difference in energy between the individual energies for placing Ge atoms at sites $i$ and $j$ (i.e. $\Delta E(\vec{r}_i)$ and $\Delta E(\vec{r}_j)$ from above) and the energy found when both sites are occupied in a periodic calculation. The perturbation energy is given by:
\begin{equation}
 \Delta E(\vec{r}_i,\vec{r}_j)= E_{ij}-\Delta E(\vec{r}_i)-\Delta E(\vec{r}_j)-E_0
\end{equation} where $E_{ij}$ is the lattice energy of structures with two Ge/Si substitutions.
\item Idem for trios and quartets of atoms.
\begin{align}
 \Delta E(\vec{r}_i,\vec{r}_j,\vec{r}_k) &=E_{ijk}-\Delta E(\vec{r}_i,\vec{r}_j)-\Delta E(\vec{r}_i,\vec{r}_k)-\Delta E(\vec{r}_j,\vec{r}_k)-\notag \\
   \qquad &-\Delta E(\vec{r}_i)-\Delta E(\vec{r}_j)-\Delta E(\vec{r}_k)-E_0  \\
 \Delta E(\vec{r}_i,\vec{r}_j,\vec{r}_k,\vec{r}_l) &= E_{ijkl}-\Delta E(\vec{r}_i,\vec{r}_j,\vec{r}_k)-\{\ldots\}^{ijk}-\Delta E(\vec{r}_j,\vec{r}_k,\vec{r}_l)- \notag \\ 
   \qquad &-\Delta E(\vec{r}_i,\vec{r}_j)-\{\ldots\}^{ijkl}-\Delta E(\vec{r}_k,\vec{r}_l)-\notag \\
   \qquad &-\Delta E(\vec{r}_i)-\Delta E(\vec{r}_j)-\Delta E(\vec{r}_k) - \Delta E(\vec{r}_l) -E_0
\end{align}

\end{enumerate} where $\{\ldots\}^{ijk}$ and $\{\ldots\}^{ijkl}$ represent all the summation terms, which are combinations of the $ijk$ and $ijkl$ indices, respectively.

Then, an effective approximate lattice energy of $N$ Si/Ge substitutions is being calculated as:
\begin{align}
\label{eq:EH_1st}
\mathcal{H} &=E_0+\sum_{i}{\Delta E(\vec{r}_i)}+\sum_{ij}{\Delta E(\vec{r}_i,\vec{r}_j)}+\sum_{ijk}{\Delta E(\vec{r}_i,\vec{r}_j,\vec{r}_k)}+ \notag\\
\qquad &+\sum_{ijkl}{\Delta E(\vec{r}_i,\vec{r}_j,\vec{r}_k,\vec{r}_l)}+\mathcal{O}(\vec{r}^N)
\end{align} where $i,j,k,l$-indexes run on the total number of configurations. Is useful to adapt the Equation \ref{eq:EH_1st} with a \textit{tensor} notation using the Einstein summation convention:
\begin{equation}
\label{eq:EH_2nd}
\mathcal{H}(N)=E_0+\epsilon_iS^i+\rho_{ij}S^iS^j+\theta_{ijk}S^iS^jS^k+\phi_{ijkl}S^iS^jS^kS^l
\end{equation}
where $S^i$ are spin-type variables wich with $1$ or $0$ represent the presence or absence, respectively, of Ge atom in the crystallographic $i$--position, $\epsilon:=\{\Delta E(\vec{r}_i)\}$, $\rho:=\{\Delta E(\vec{r}_i,\vec{r}_j)\}$, $\theta:=\{\Delta E(\vec{r}_i,\vec{r}_j,\vec{r}_k)\}$ and $\phi:=\{\Delta E(\vec{r}_i,\vec{r}_j,\vec{r}_k,\vec{r}_l)\}$. We can readapt the Equation \ref{eq:EH_2nd} to sum on the inequivalent configurations using a dictionary, $\delta$, which connects each configuration with the calculated equivalent configuration.
\begin{align}
\label{eq:EH}
\mathcal{H}(N)&=E_0+\epsilon_\alpha\delta_i^\alpha S^i+\frac{\rho_{\alpha\beta}}{N-1}\delta^{\alpha\beta}_{ij}S^iS^j +
                \frac{2\theta_{\alpha\beta\gamma}}{(N-2)(N-3)}\delta^{\alpha\beta\gamma}_{ijk}S^iS^jS^k + \notag\\
       \qquad &+\frac{2\phi_{\alpha\beta\gamma\zeta}}{(N-3)(N-4)}\delta^{\alpha\beta\gamma\zeta}_{ijkl}S^iS^jS^kS^l
\end{align}

\subsection{Results}
\begin{table} 	  
   \caption{Summary of synthesis results at 175 $^\circ$C}	  
   \label{tbl:synth}	  
   \centering
   \footnotesize
   \begin{tabular}{cccc}
   \hline	 
    Ge$_f$ & Time (hours) & Yield (wt. \%)  & Phase\textsuperscript{\emph{a}}  \\
   \hline    
  0  &  27  &  31.1   &   amorphous (+ HPM-1) \\
  0  &  48  &  21.2   &  HPM-1 \\ 
  0  &  143  &  29.7  &  HPM-1 \\ 
  0  &  264  &  29.6  &  HPM-1 \\ 

  0.009  &  25  &  22.5  &  HPM-1 + amorphous  \\
  0.009  &  48  &  23.7  &  HPM-1 \\ 
  0.009  &  144  &  23.7  &  HPM-1 \\ 
  0.009  &  240  &  24.1  &  HPM-1 \\ 

  0.019  &  25  &  25.1  &  HPM-1 \\ 
  0.019  &  48  &  24.1  &  HPM-1 \\ 
  0.019  &  144  &  25.4  &  HPM-1 \\ 
  0.019  &  240  &  25.4  &  HPM-1 \\ 

  0.032  &  25  &  27.1  &  HPM-1 \\ 
  0.032  &  48  &  27.9  &  HPM-1 \\ 
  0.032  &  144  &  26.16  &  HPM-1 \\ 
  0.032  &  240  &  26.3  &  HPM-1 \\ 

  0.09  &  25  &  33.1 &  HPM-1 \\ 
  0.09  &  48  &  31.4  &  HPM-1 \\ 
  0.09  &  144  &  32.2  &  HPM-1 \\ 
  0.09  &  240  &  34.4  &  HPM-1 \\ 

  0.166  &  25  &  34.4  &  HPM-1 \\ 
  0.166  &  48  &  34.5  &  HPM-1 \\ 
  0.166  &  144  &  33.3  &  HPM-1 \\ 
  0.166  &  237  &  36.1  &  HPM-1 \\ 

  0.2  &  25  &  36.4  &  HPM-1 \\ 
  0.2  &  48  &  35.7  &  HPM-1 \\ 
  0.2  &  144  &  36.1  &  HPM-1 \\ 
  0.2  &  240  &  35.5  &  HPM-1 \\ 

  0.4  &  25  &  39.3  &  HPM-1 \\ 
  0.4  &  48  &  39.7  &  HPM-1 \\ 
  0.4  &  144  &  38.0  &  HPM-1 \\ 
  0.4  &  240  &  37.9  &  HPM-1 \\ 

  0.6  &  25  &  24.9  &  HPM-1 \\ 
  0.6  &  48  &  26.8  &  HPM-1 \\ 
  0.6  &  144  &  27.6  &  HPM-1 \\ 
  0.6  &  240  &  32.5  &  HPM-1 \\ 

  0.8  &  25  &  25.4  &  HPM-1 \\ 
  0.8  &  48  &  23.3  &  HPM-1 \\ 
  0.8  &  144  &  26.6  &  HPM-1 \\ 

  0.9  &  25  &  5.5 &  HPM-1 \\ 
  0.9  &  48  &  18.1  &  HPM-1 \\ 
  0.9  &  144  &  21.8  &  HPM-1 \\ 

  1  &  27  &  0  & -  \textsuperscript{\emph{b}} \\
  1  &  113  &  1.7&  HPM-1 \\ 
  1  &  200  &  2.6  &  Q+Arg (+HPM-1)\textsuperscript{\emph{c}} \\
  1  &  96  &  10.4  &  HPM-1 (+Q)\textsuperscript{\emph{c,d}}\\ 
  1  &  102  &  9.5  &  HPM-1 +Q\textsuperscript{\emph{c,d}})\\ 
 \hline  
\end{tabular}\\
\textsuperscript{\emph{a}} Major phases are listed first, very minor phases appear between parentheses. 
\textsuperscript{\emph{b}} No solids could be collected.
\textsuperscript{\emph{c}} Q is the Quartz-like and Arg is the Argutite-like \ce{GeO2} phases.
\textsuperscript{\emph{d}} The last two entries correspond to a different synthesis run in the same nominal conditions as the preceeding ones.
\end{table}

\begin{figure}
	\includegraphics[width=\textwidth]{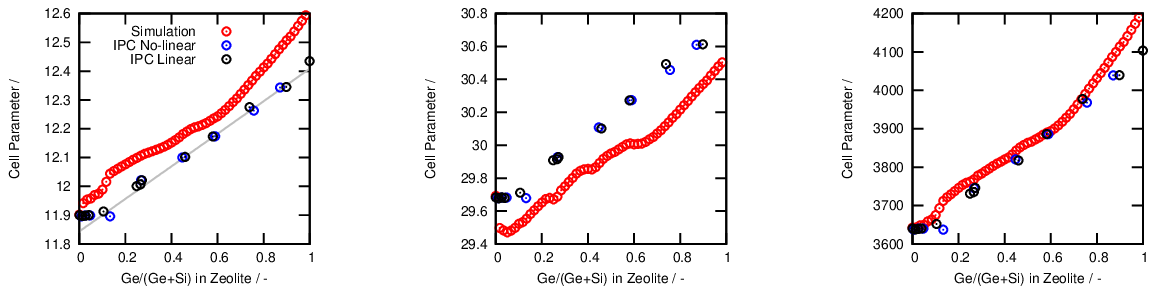}
	\caption{Cell parameters vs Ge$_f$ computed with the EH.}
	\label{fgr:cell}
\end{figure}

\begin{figure}
 \includegraphics[width=0.6\textwidth]{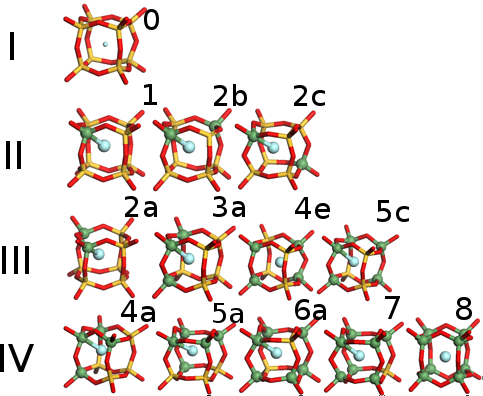}
 \caption{ Some of the energy-minimised D4Rs configurations. They are classified according to the resonance. Some structural distortions are distinguishable with respect to the configuration of pure silica.}
\label{fgr:D4R_types}
\end{figure}

\begin{figure}[!h]
	\includegraphics[width=0.6\textwidth]{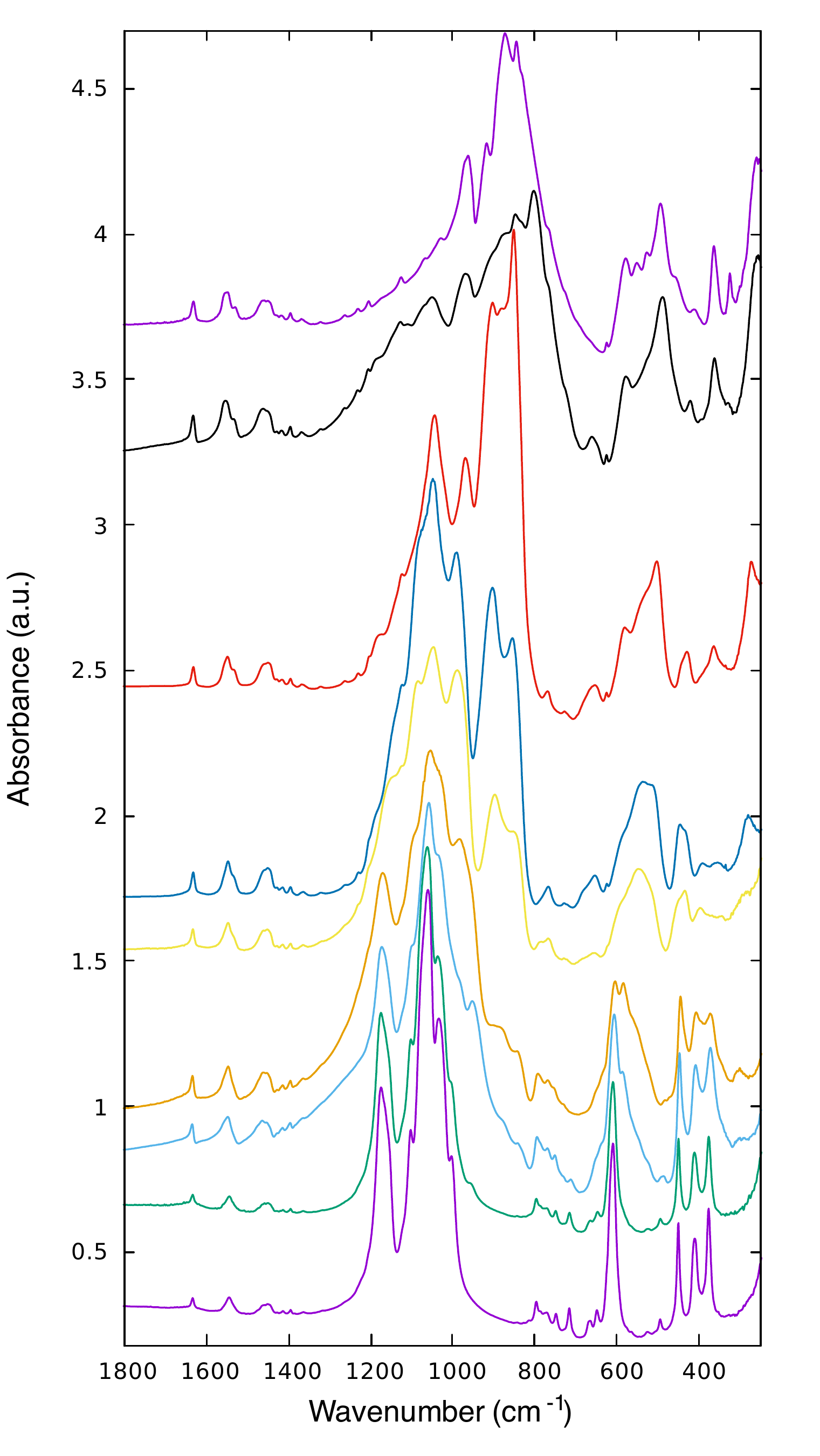}
		\caption{Infrared species of STW zeolites obtained from gels with Ge$_f$ = 0, 0.01, 0.09, 0.17, 0.4, 0.6, 0.8, 0.9 and 1 (from bottom to top)}
    \label{fgr:ir}
\end{figure}

\begin{figure}[!h]
	\includegraphics[width=0.9\textwidth]{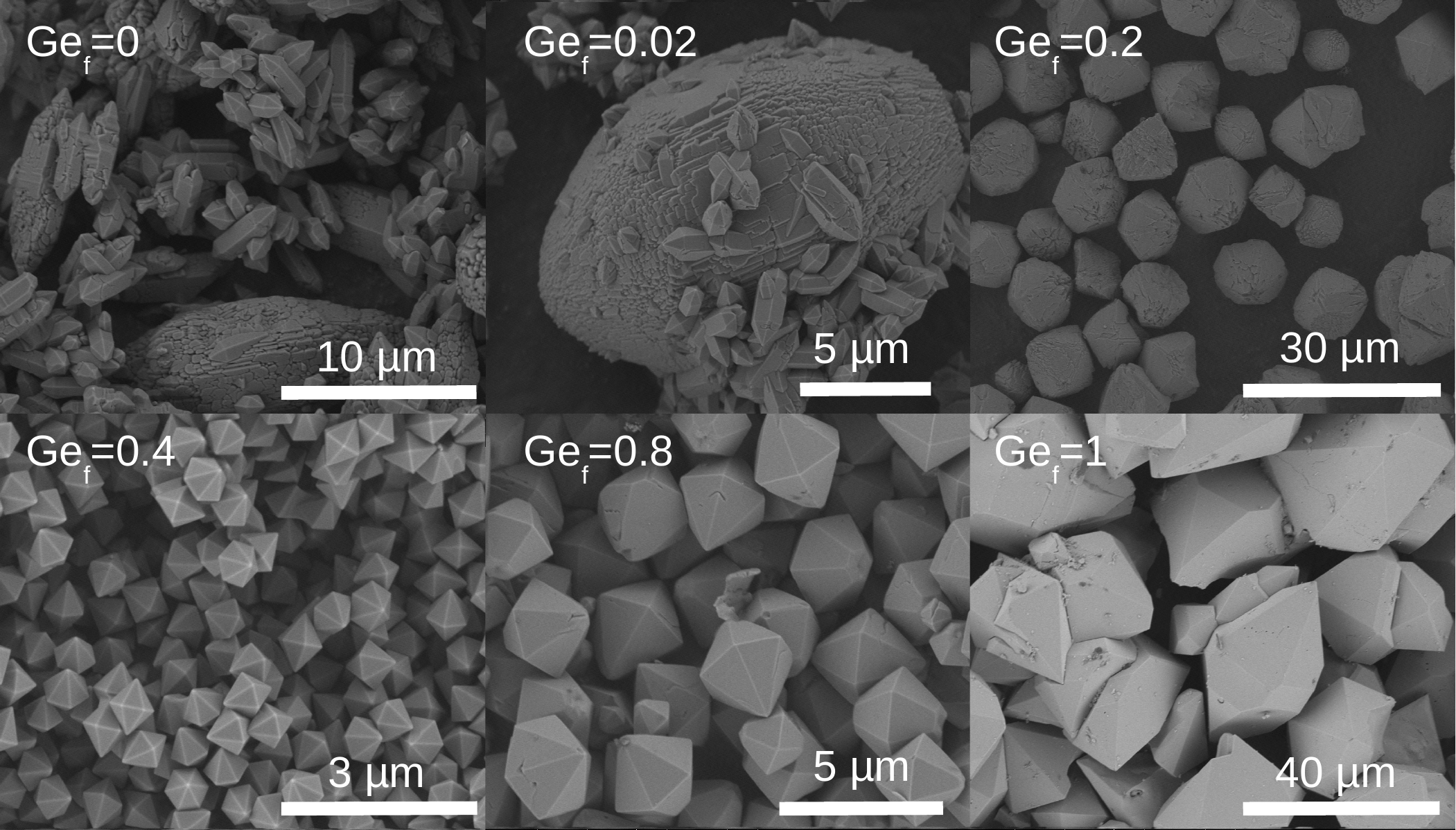}
		\caption{FESEM images of HPM-1 zeolites prepared at different Ge$_f$ levels.}
    \label{fgr:fesem}
\end{figure}

\begin{figure}[!h]
	\includegraphics[width=0.7\textwidth]{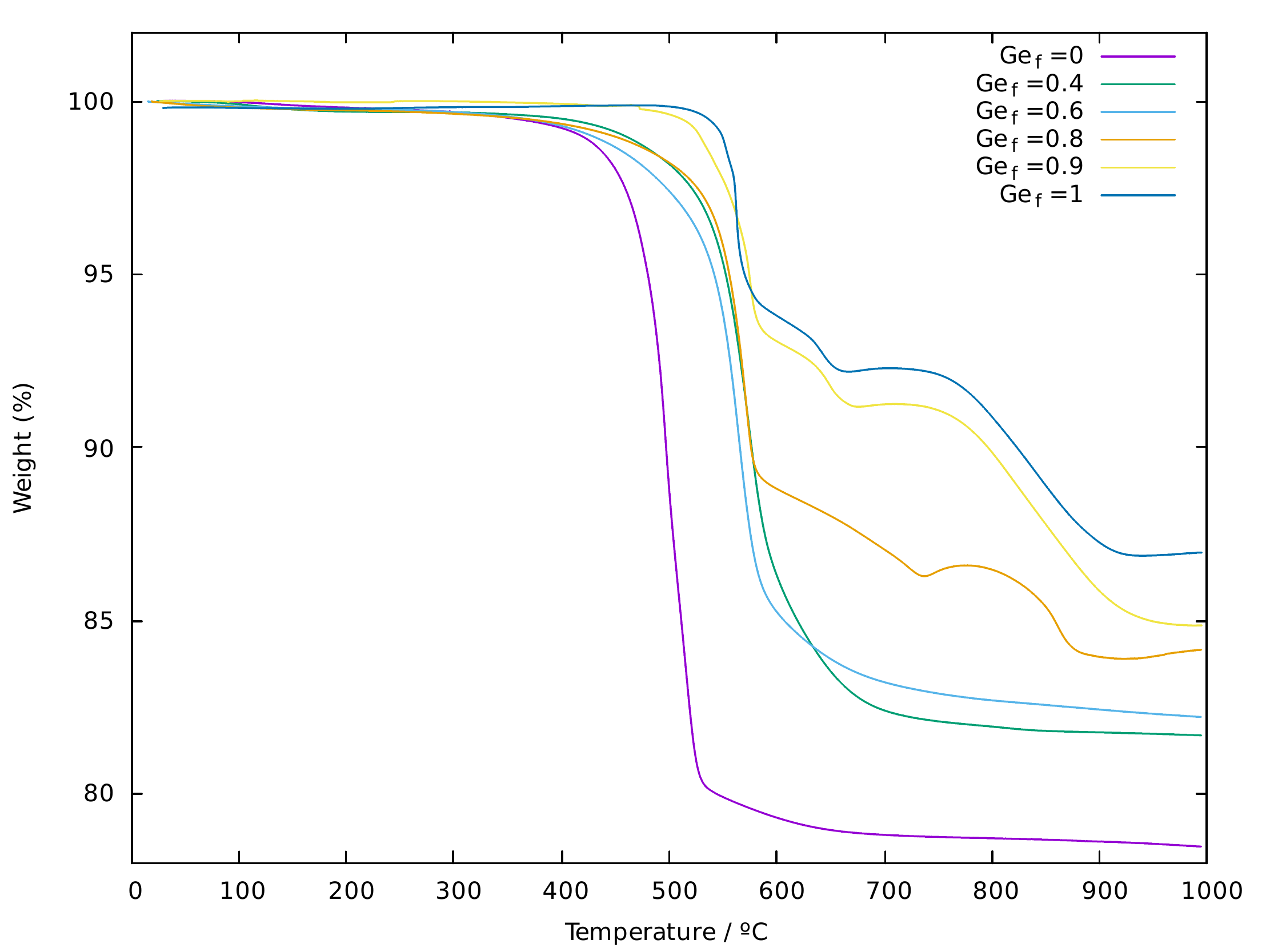}
		\caption{Thermograms of HPM-1 zeolites prepared at different Ge$_f$ levels.}
    \label{fgr:tg}
\end{figure}

\begin{figure}[h!]
	\includegraphics[width=0.65\textwidth]{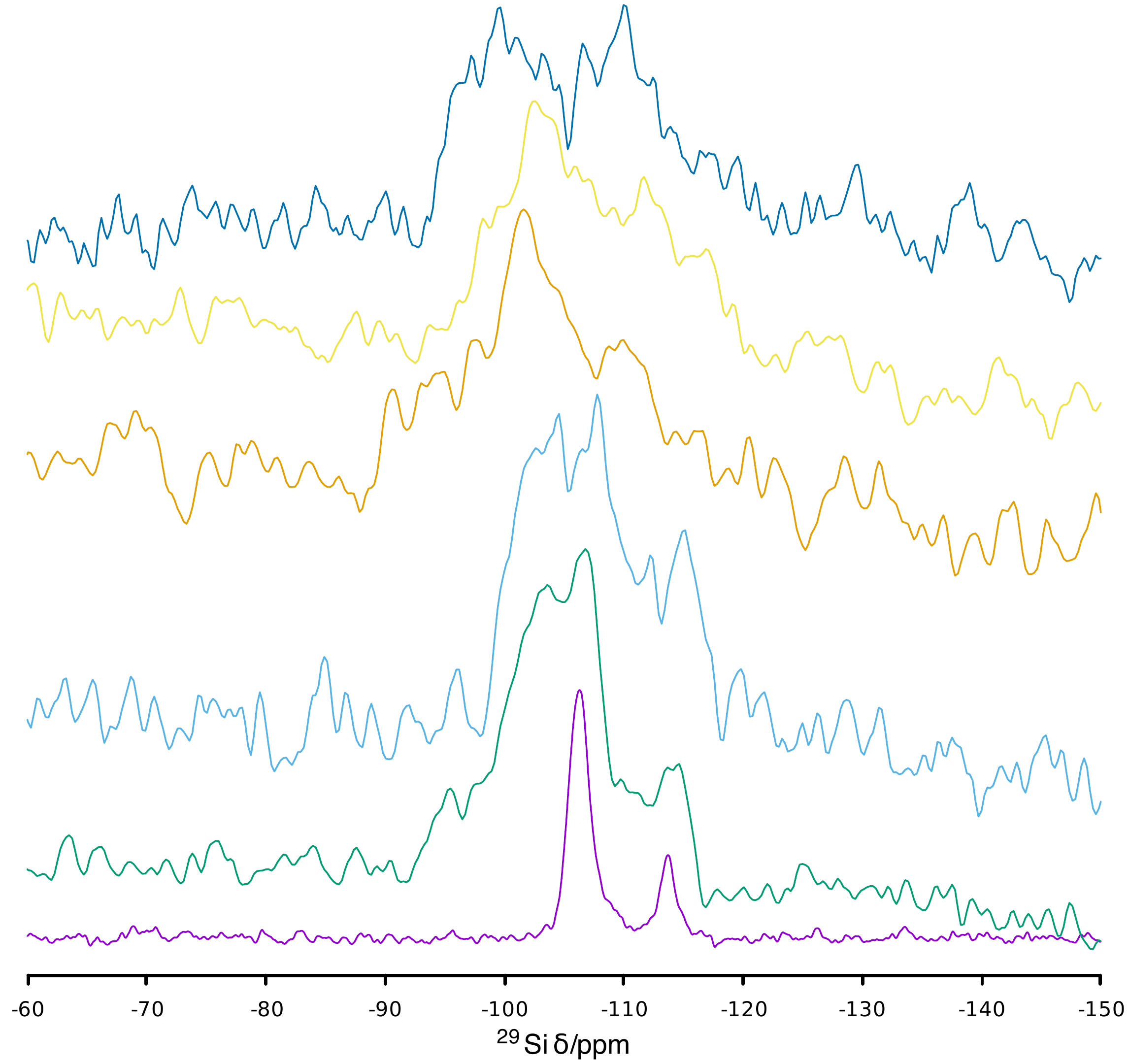}
		\caption{Direct irradiation $^{29}$Si MAS spectra of (Ge,Si)-STW with Ge$_f$= 0.00, 0.166, 0.20, 0.40, 0.60 and 0.80(from bottom to top). For every spectra 2048 scans were acquired. Recycle delays are 60s for the pure silica sample and 180s for the rest.}
    \label{fgr:29sidirecto}
\end{figure}

\begin{figure}[h!]
	\includegraphics[width=0.5\textwidth]{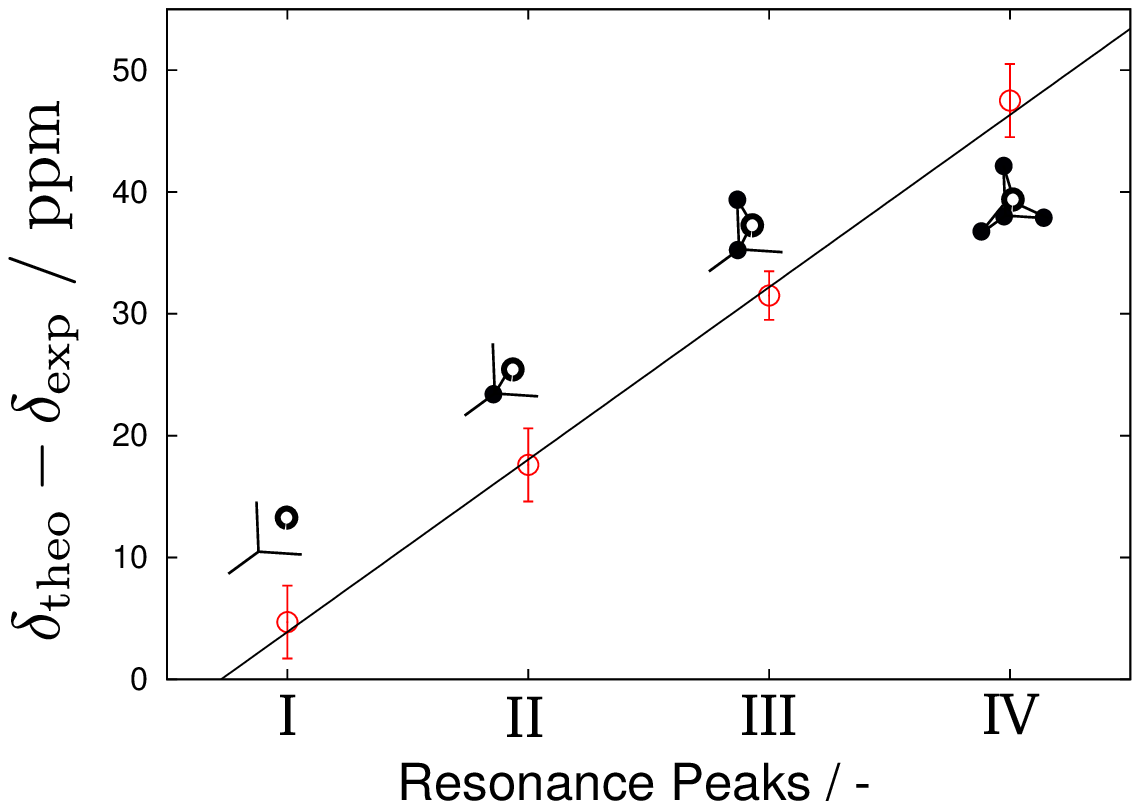}
		\caption{Difference between calculated and experimental chemical shifts for each $^{19}F$ resonance according to our assignment (see main text).}
    \label{fgr:localenvironments}
\end{figure}

\clearpage
\subsection{Rietveld details}
\begin{figure}
  \centering
  \makeatletter 
  \renewcommand{\thefigure}{S\@arabic\c@figure}
  \makeatother
  \includegraphics[width=4.5in]{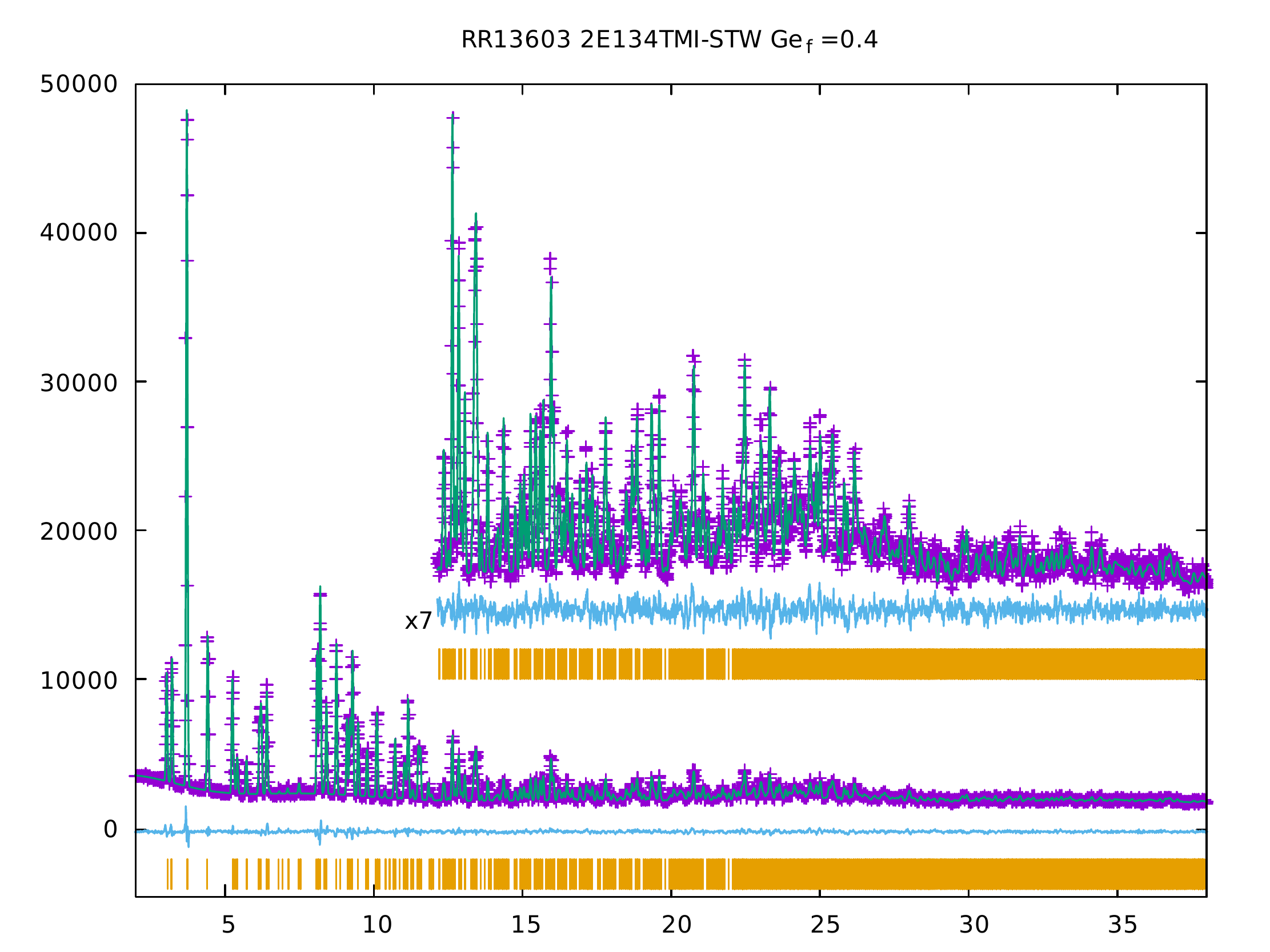}
  \caption{Observed (+) and calculated (solid line) powder X-ray diffractograms for as-made Ge,Si-HPM-1 with $Ge_f=0.4$ refined in space group $P6_122$. Vertical tic marks indicate the positions of allowed reflections. The lower trace is the difference plot. $\lambda$=0.56383 \r{A}.}
  \label{fgr:riet0.4}
\end{figure}

\begin{figure}
  \centering
  \makeatletter 
  \renewcommand{\thefigure}{S\@arabic\c@figure}
  \makeatother
  \includegraphics[width=4.5in]{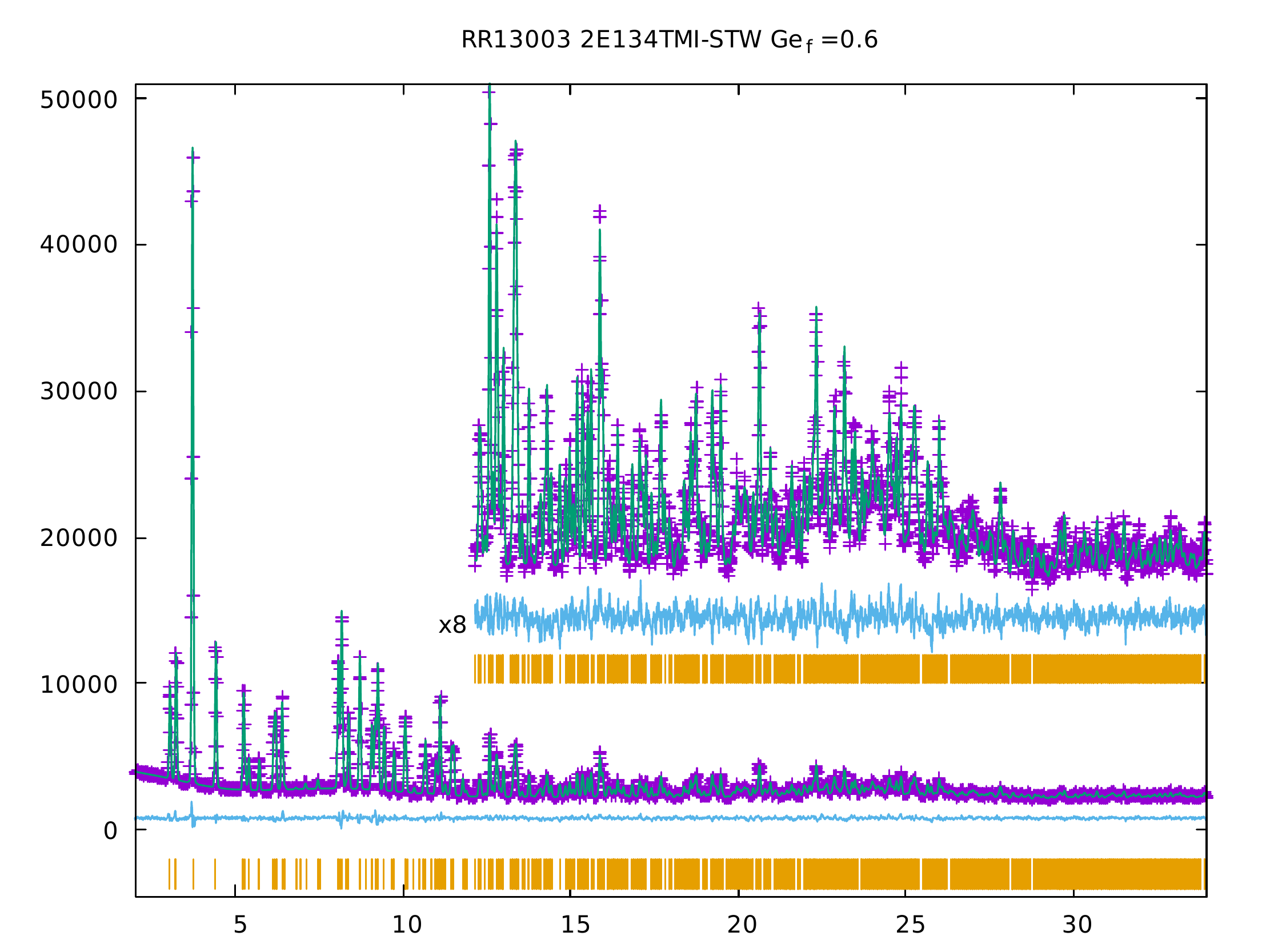}
  \caption{Observed (+) and calculated (solid line) powder X-ray diffractograms for as-made Ge,Si-HPM-1 with $Ge_f=0.6$ refined in space group $P6_122$. Vertical tic marks indicate the positions of allowed reflections. The lower trace is the difference plot. $\lambda$=0.56383 \r{A}.}
  \label{fgr:riet0.6}
\end{figure}

\begin{figure}
  \centering
  \makeatletter 
  \renewcommand{\thefigure}{S\@arabic\c@figure}
  \makeatother
  \includegraphics[width=4.5in]{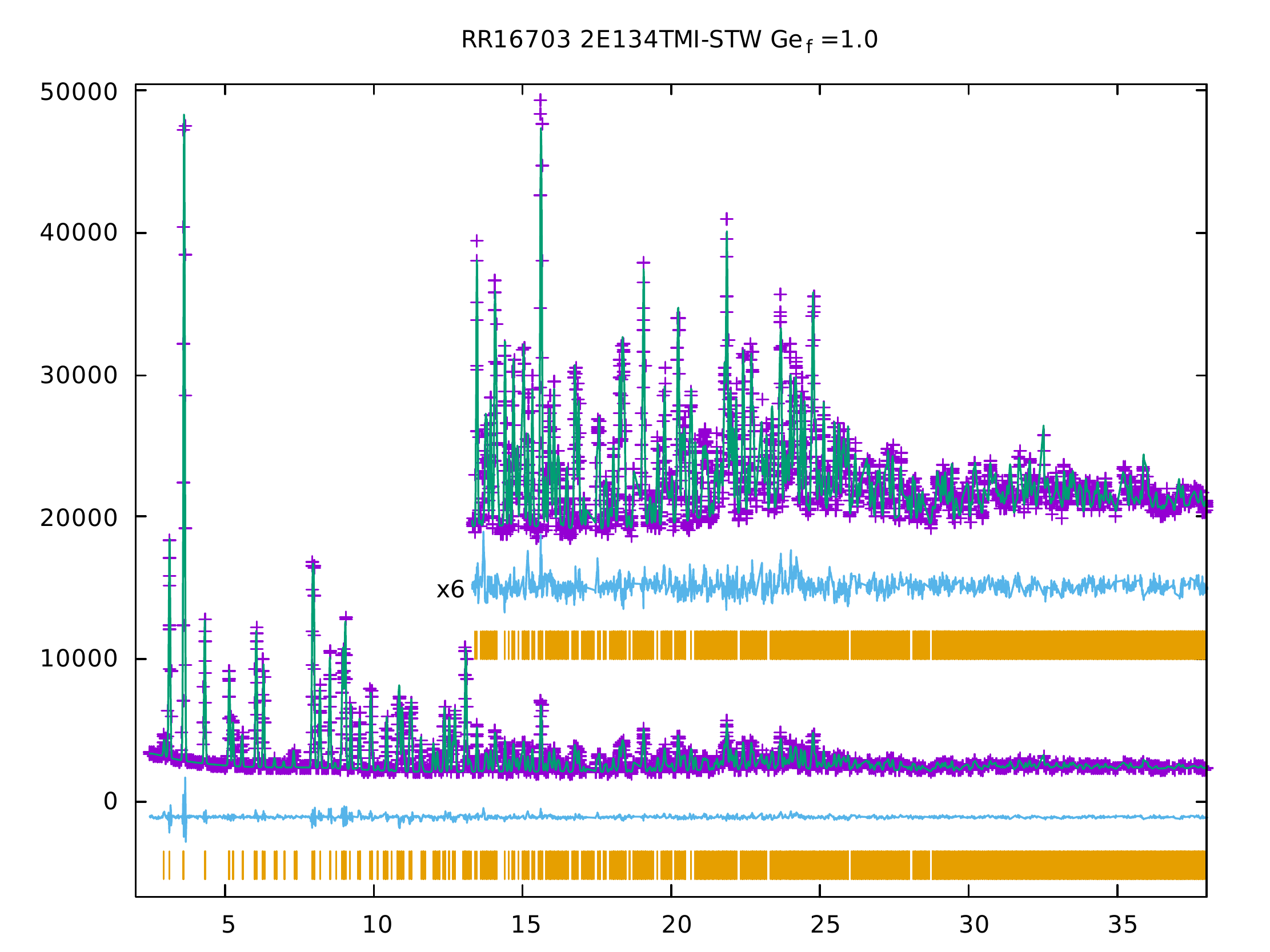}
  \caption{Observed (+) and calculated (solid line) powder X-ray diffractograms for as-made Ge-HPM-1 ($Ge_f=1.0$) refined in space group $P6_122$. Vertical tic marks indicate the positions of allowed reflections. The lower trace is the difference plot. $\lambda$=0.56383 \r{A}.}
  \label{fgr:riet1}
\end{figure}

The starting model for Rietveld refinement of samples prepared with Ge$_f$=0.4, 0.6 and 1 was the refined structure of pure silica HPM-1,\cite{Rojas2013b} in space group $P6_122$ modified to have the unit cell dimensions determined from conventional powder XRD data and a Ge occupation of all crystallographic positions initially set at 0.4, 0.6 and 1.0, respectively (Si occupancies of 0.6, 0.4 and 0.0, respectively). Although the sample with Ge$_f=1$ appeared as phase-pure in the conventional XRD pattern, synchrotron radiation showed the presence of small traces of quartz-like \ce{GeO2}, and the corresponding regions were excluded from the refinement. A Lobanov and Alte da Veiga absorption correction was applied.\cite{Larson2004} Scale factor, unit cell and profile parameters were refined, with a shifted Chebyschev function initially with 16 fixed parameters to simulate the background. Then, the Ge, Si, O and F atoms were allowed to move, initially with soft restrains on T-O and O-O distances and with Ge and Si in each crystallographic site constrained to move together. Then, the position and orientation of the organic SDA was refined as a rigid body consisting of the imidazolium ring with the three methyl substituents as a rigid unit plus the ethyl group as a satellite that could freely rotate along the C2-C9 bond. The hydrogen atoms were omitted but the fractional occupancies of the C atoms were adjusted to account for the electrons of the bonded H. The weight of the distance restraints was gradually reduced and eventually eliminated. In the final stages of the refinements, atom displacement factors (grouped by atom type), background and fractional occupancies of Ge and Si in each crystallographic site (constrained to amount to a full occupancy of each site) were included in the refinement. Final crystallographic data are summarized in Tables \ref{tbl:riet}, the final Rietveld plots are given in Figures \ref{fgr:riet0.4}, \ref{fgr:riet0.6} and \ref{fgr:riet1} 
 and the corresponding cif files are provided as supplementary material.

\begin{table*}[h]  
  \centering
  \makeatletter 
   \renewcommand{\thetable}{S\@arabic\c@table}
   \renewcommand\@biblabel[1]{#1}    
   \makeatother
   \caption{Crystallographic and Experimental Parameters for the Rietveld Refinement of as-made Ge,Si-HPM-1 phases (wavelength: 0.56383 \r{A}, Temperature 293K)}
  \label{tbl:riet}
  \begin{tabular}{cccc}
    \hline
     Nominal $Ge_f$ & 0.4                  & 0.6   	&1\\
     Refined $Ge_f$ & 0.396                &0.574	& -\\
     2$\theta$ range & 2.00-38.32                &                   2.00-33.97       & 2.50-44.9\\
     no. of data points & 3733                   &                3198                & 4241\\
     no. of reflections & 1327                   &                977                 & 1792\\
     Space Group & $P6_122$                      &                $P6_122$            & $P6_122$ \\
     unit cell parameters (\r{A}) &  &&\\              
     \hspace{20pt} $a,b$ & 12.09289(16)          &                  12.16714(17)     & 12.42671(11)\\
     \hspace{20pt} $c$ & 30.0839(5)              &                  30.2519(6)       & 30.6310(5)\\
     Cell volume (\r{A}$^3$) &  3810.01(12)      &                  3878.46(13)      &  4096.41(8)\\
     Residuals&&&\\                              
     \hspace{20pt} R$_{wp}$ & 3.06\%             &                   2.82\%          & 3.66\%\\
     \hspace{20pt} R$_p$& 2.36\%                 &                   2.23\%          &2.65\%\\
     \hspace{20pt} R$_{F^2}$& 7.76\%             &                   7.53\%          & 10.58\%\\
     \hspace{20pt} reduced $\chi^2$ & 2.513      &                   2.527            & 3.98\\
    \hline  
  \end{tabular}

\end{table*}

\begin{table*}[h]  
  \centering
  \makeatletter 
   \renewcommand{\thetable}{S\@arabic\c@table}
   \renewcommand\@biblabel[1]{#1}    
   \makeatother
   \caption{Average bond distances and angles in Ge,Si-HPM-1 phases}
  \label{tbl:disagl}
  \begin{tabular}{cccc}
    \hline
   Average distance  (\AA) &  Ge$_f=0.4$ & Ge$_f=0.6$ & Ge$_f=1$\\
    \hline
   T1-O&  1.654 & 1.662 & 1.668\\
	T2-O & 1.656 &1.670 & 1.714\\
 	T3-O & 1.637 &1.647 & 1.706\\
	T4-O & 1.667 &1.689  & 1.724\\
	T5-O & 1.635 &1.642 & 1.724\\
   T1-F & 2.57 &2.63 & 2.57\\
	T2-F & 2.79 &2.76  & 2.89\\
	T3-F & 2.70 & 2.71 & 2.72\\
	T4-F & 2.74 &2.75 & 2.75\\
    \hline
   Average angle ($^\circ$)&&&\\
   O-T1-O & 109.3 & 109.2 & 109.2\\
	O-T2-O & 109.4 &109.4 & 109.3\\
   O-T3-O & 109.1 &109.0 & 109.4\\
   O-T4-O & 109.4 &109.4 & 109.2\\
   O-T5-O & 109.4 & 109.4 & 109.6\\
    \hline  
  \end{tabular}
\end{table*}

\clearpage

\bibliography{biblio}
\end{document}